\documentclass[twocolumn]{aa}

\usepackage{graphicx}
\usepackage{hyperref}
\usepackage{dblfloatfix}
\usepackage{placeins}
\usepackage{txfonts}
\begin{document} 

   \title{Stellar halos of bright central galaxies II}
   \subtitle{Scaling relations, colours and metallicity evolution with redshift}

   \author{Emanuele Contini \inst{1} 
   \thanks{\email{emanuele.contini82@gmail.com, yi@yonsei.ac.kr}}
          \and Marilena Spavone  \inst{2}
          \and Rossella Ragusa \inst{2}
          \and Enrichetta Iodice \inst{2}
          \and Sukyoung K. Yi \inst{1}
          }
           \institute{Department of Astronomy and Yonsei University Observatory, Yonsei University, 50 Yonsei-ro, Seodaemun-gu, Seoul 03722, Republic of Korea
            \and
          INAF Osservatorio Astronomico di Capodimonte, Salita Moiariello 16, 80131 Napoli, Italy}

\abstract
{}
{We investigate the formation and evolution of stellar halos (SHs) around bright central galaxies (BCGs), focusing on their scaling relations, colours, and metallicities across cosmic time, and we compare model predictions with ultra-deep imaging data.}
{We use the semi--analytic model \textsc{FEGA25}, applied to merger trees from high-resolution dark matter simulations, including an updated treatment of intracluster light (ICL) formation. SHs are defined as the stellar component within a physically motivated transition radius, linked to the structural properties of the host halo. Predictions are compared with observations from the VST Early-type GAlaxy Survey (VEGAS) and Fornax Deep Survey (FDS).}
{The SH mass correlates well with both BCG and ICL masses, with tighter scatter in the SH--ICL relation. The transition radius peaks at 30--40 kpc nearly independent of redshift in the model predictions, but can reach $\sim400$ kpc in the most massive halos, after $z=0.5$. SHs and ICL show nearly identical colour distributions at all epochs, both reddening toward $z=0$. At $z=2$, SHs and the ICL are $\sim0.4$ dex more metal--poor than BCGs, but the gap shrinks to $\sim0.1$ dex by the present time. Observed colours are consistent with model predictions, while observed metallicities are lower, suggesting a larger contribution from disrupted dwarfs.}
{SHs emerge as transition regions between BCGs and the ICL, dynamically and chemically coupled to both. Their properties depend on halo concentration, ICL formation efficiency, and the progenitor mass spectrum. Upcoming wide--field photometric and spectroscopic surveys (e.g. LSST, WEAVE, 4MOST) will provide crucial tests by mapping structure, metallicity, and kinematics in large galaxy samples.}


   \maketitle
%

\section{Introduction}\label{sec:intro}

Stellar halos (SHs) are diffuse, low surface–brightness (LSB) stellar envelopes that surround galaxies and preserve a fossil record of past accretion events and early phases of in-situ star formation \citep{bullock2005,cooper2010}. Their study has become a cornerstone for understanding the hierarchical build-up of galaxies within the $\Lambda$CDM framework, since halo stars are long-lived and retain chemical and dynamical imprints of their progenitors \citep{helmi2008,font2011,duc2015,iodice2016,lane2022,beltrand2024}.

Photometric surveys of resolved red-giant branch (RGB) stars in nearby galaxies have shown that colours can serve as robust proxies for metallicity, under reasonable assumptions about stellar ages \citep{monachesi2016,harmsen2017}. These surveys reveal a remarkable diversity: while some halos display strong negative colour and metallicity gradients (bluer, more metal-poor populations at larger radii), others show almost flat profiles over tens of kiloparsecs \citep{mouhcine2005,ibata2014}. In the Milky Way, spectroscopic surveys have established a clear dichotomy between the inner halo, relatively metal-rich, and the outer halo, more metal-poor, consistent with a two-component formation scenario \citep{carollo2007,beers2012,deason2014,helmi2018}. Similarly, M31 exhibits a pronounced metallicity gradient extending to large radii, suggesting that its halo has been significantly shaped by the accretion of relatively massive satellites \citep{kalirai2006,gilbert2014}.

A key emerging result is that SH metallicities correlate with host galaxy stellar mass or luminosity: more massive galaxies tend to host more metal-rich halos, a relation consistent with expectations from the mass–metallicity relation of dwarf satellites \citep{mouhcine2005,harmsen2017,dsouza2018}. However, there is large halo-to-halo scatter, reflecting the stochasticity of hierarchical assembly and the diversity of accretion histories \citep{merritt2016,amorisco2017}. Chemical abundance ratios, particularly [$\alpha$/Fe], provide further constraints: high [$\alpha$/Fe] ratios trace rapid star formation in massive progenitors or in-situ components, while lower values indicate extended star formation in low-mass dwarfs \citep{venn2004,deason2016}.

Recently, ultra-deep wide-field photometric surveys have extended SH studies to a larger variety of environments, including galaxy groups and clusters. The VST Early-type GAlaxy Survey (VEGAS) survey has revealed extended halos, tidal features, and intracluster light (ICL) in several nearby systems \citep{capaccioli2015,iodice2017,iodice2019,spavone2017,spavone2018,spavone2020,spavone2021,spavone2022,spavone2024,ragusa2023}. These studies highlight the strong connection between the outer envelopes of massive early-type galaxies (ETGs) and their accretion histories, with signatures of past mergers, stripping and disruption events. Such results complement stellar population studies in the Local Group, extending SH science to a statistical sample of galaxies in diverse environments.

Theoretical models and cosmological simulations have been instrumental in interpreting these observations. Early semianalytic and particle-tagging models highlighted the role of a few dominant progenitors in building up halos \citep{bullock2005,cooper2010}. More recent hydrodynamical simulations, such as AURIGA \citep{grand2017}, FIRE \citep{hopkins2014}, and IllustrisTNG \citep{nelson2018}, demonstrate that both accreted and in-situ components are important: in-situ stars are typically more metal-rich and centrally concentrated, producing steeper gradients, whereas accreted halos are flatter unless shaped by the disruption of massive satellites \citep{font2011,pillepich2018,monachesi2019,elias2020}. In summary, studies of SH colours and metallicities reveal a complex and diverse picture. Observations point to correlations with host mass and stochastic variations tied to accretion history, while simulations stress the interplay between in-situ star formation and accreted material (e.g., \citealt{cooper2015,wright2024}). Future wide-field imaging surveys and spectroscopic campaigns—such as LSST \citep{ivezic2019}, WEAVE \citep{jin2024}, and 4MOST \citep{dejong2019}—will expand halo samples to statistically significant sizes, enabling robust comparisons with cosmological models and yielding new insights into galaxy formation and evolution \citep{cooper2011,helmi2020,spavone2022}.

Recent simulation suites have substantially expanded both statistics and physical realism for SH studies. The augmented AURIGA Project adds 40 new Milky Way–mass zooms (plus 26 dwarf-mass systems) at fixed physics, enabling stronger constraints on halo-to-halo scatter in ex-situ fractions, metallicity gradients, and their connection to merger histories \citep{grand2017}. Within AURIGA, new analyses connect the presence of strongly radially anisotropic debris and steep negative [Fe/H] gradients to the dominance of one or a few massive accretion events, refining and extending earlier gradient–assembly correlations (e.g. “Gaia–Sausage–Enceladus”-like progenitors, \citealt{belokurov2018,helmi2018}).

In the cluster regime, the TNG-Cluster project \citep{nelson2024} opens the door to statistical studies of BCGs+SHs+ICL within the same galaxy-formation model as IllustrisTNG. First results map stellar/ICM properties and assembly histories, providing a framework to trace ex-situ build-up and outer-envelope growth at high masses. Complementary large-volume and comparison projects strengthen these trends. In The Three Hundred hydrodynamical simulations \citep{cui2018}, the ICL fraction correlates with the cluster dynamical state, supporting a picture where late-time accretion and tidal processing grow diffuse outer components—an inference directly relevant to massive-galaxy SHs and their metallicity gradients in dense environments \citep{contreras-santos2024}. At group/cluster scales in TNG300 \citep{nelson2019}, recent work revisits BCG growth and the timing of ex-situ assembly, highlighting late-time outer-envelope build-up and providing predictions for metallicity and age trends with radius that can be confronted with deep imaging and spectroscopy \citep{montenegro-taborda2023}.

Finally, related zoom studies continue to clarify the in-situ vs. accreted imprint on chemo-kinematics. Edge-on maps in AURIGA connecting age–metallicity–[$\alpha$/Fe] to merger-driven heating offer templates for interpreting thick-disc/inner-halo overlaps and their projected gradients. These maps emphasize that spherically averaged metallicity profiles are intrinsically steeper than minor-axis projections typically measured observationally \citep{pinna2024}.

In this paper, we expand upon the analysis presented in \citet{contini2024c}, where we first implemented SH formation in our semianalytic model of galaxy formation. Here, we investigate the main scaling relations between SH and BCG–ICL mass, and between SH mass and the transition radius in the first part, followed by an analysis of their colours and metallicities in the second part, as a function of redshift from $z=2$ down to the present time.

The remainder of the paper is structured as follows. In Section~\ref{sec:methods}, we describe the implementation of SH formation in our model and provide details about the simulations and the accompanying observational data. In Section~\ref{sec:results}, we present our analysis, highlight and discuss the main results. Finally, in Section~\ref{sec:conclusions}, we summarize the key conclusions. Unless otherwise stated, stellar and halo masses are corrected for $h=0.68$, and stellar masses are derived assuming a Chabrier initial mass function \citep{chabrier2003}.

\section{Methods}\label{sec:methods}
For the analysis that follows in Section~\ref{sec:results}, we take advantage of the state-of-the-art semianalytic model {\small FEGA25} (Formation and Evolution of GAlaxies), in its latest version as detailed in \citet{contini2025a}, specifically with the {\small AGNeject1} option for the hot gas ejection mode driven by active galactic nuclei (AGN) feedback (see \citealt{contini2025b} for further details). {\small FEGA25} implements updated prescriptions for baryonic physics, including positive AGN feedback, a renewed supernova (SN) feedback scheme, and a more sophisticated star formation law \citep{contini2024c}.

\subsection{Semi-analytic model and ICL formation}
\label{subsec:fega_icl}
Since the key focus of this work is the redshift evolution of the main properties of SHs, we begin with a detailed description of their formation. Because SHs originate from the diffuse light, or ICL, we first provide a brief summary of its formation. In {\small FEGA25}, the ICL arises through several channels, whose relative importance varies across its assembly history. The three main channels are stellar stripping, mergers, and pre-processing. Stellar stripping of satellite galaxies contributes a significant fraction of diffuse light, as stars are removed from satellites due to the potential well of the host halo. In some cases, albeit less common, tidal forces can be strong enough to completely disrupt a satellite galaxy.

Another important channel is represented by mergers between central and satellite galaxies, both minor and major. At each merger event, the code assigns a fraction of the satellite’s stars to the ICL component associated with the central galaxy, with an average of $20\%$ and a scatter of $\pm5\%$. The last, channel is pre-processing, which can be considered a sub-channel, since pre-processed ICL ultimately originates from either stellar stripping or mergers (or both). In short, pre-processed diffuse light forms in halos other than that of the central galaxy and is subsequently accreted when those halos merge with the host.

In \citet{contini2024}, and previously also in \citet{contini2023}, we showed that stellar stripping is the dominant channel of ICL formation, while mergers play a secondary role. However, as repeatedly discussed in earlier works, the net balance between the two channels strongly depends on the definition of mergers (see, e.g., \citealt{contini2018} for a detailed discussion, as well as \citealt{joo2025} and \citealt{gendron2025} for comparisons). We stress here that, in the model, the ICL is a physically motivated component whose mass and spatial distribution are determined by the cumulative effects of these processes along the halo assembly history, and not imposed
by construction.

\subsection{Definition of the stellar halo and transition radius}
\label{subsec:sh_definition}
In {\small FEGA25}, the SH forms directly from stars originally belonging to the ICL. The first implementation was introduced in \citet{contini2024d} through the definition of a transition region between the central galaxy (CG) and the ICL, itself defined as the stars lying within the transition radius $R_{\rm{trans}}$. This radius is linked to the halo concentration, which is computed on the basis of a Navarro-Frenk-White (NFW) dark matter distribution \citep{nfw1997}. The ICL concentration is directly tied to halo concentration via the following equation:
\begin{equation}\label{eqn:conc}
c_{\rm{ICL}} = \gamma c_{200} = \gamma \frac{R_{200}}{R_{\rm{s, DM}}} = \frac{R_{200}}{R_{\rm{s, ICL}}} , ,
\end{equation}
where $R_{200}$ and $R_{\rm{s, DM}}$ represent the virial and scale radii of the DM halo, respectively, $R_{\rm{s, ICL}}$ is the scale radius of the ICL mass distribution, and $\gamma$ is a parameter quantifying how much more concentrated the ICL is relative to the DM. This parameter was calibrated in \citet{contini2024d} using a Markov chain Monte Carlo algorithm, and typically ranges between 1 and 3.
This model assumes that the ICL distribution follows an NFW profile (see, e.g., \citealt{contini2020b} for further details), but with higher concentration. From Equation~\ref{eqn:conc}, the transition radius $R_{\rm{trans}}$ is then defined as
\begin{equation}\label{eqn:rtrans}
R_{\rm{trans}} = \frac{R_{200}}{c_{\rm{ICL}}} , ,
\end{equation}
which is essentially the analogue of the NFW scale radius for the ICL.

To model the SH, we assume that all stars belonging to the ICL and lying within the transition radius form the SH, considered as an intermediate region between the CG and the ICL. It must be noted that this implementation does not explicitly determine whether these stars are gravitationally bound to the central galaxy. Consequently, the SH can be regarded either as a distinct third component of the BCG+SH+ICL system, or as a transition zone. Stars can move between the ICL and SH components, and in certain conditions they can migrate from the SH to the BCG. By construction, the SH mass is therefore closely linked to the ICL mass and spatial distribution. In particular, a correlation between SH and ICL mass is an expected outcome of the adopted definition, while the residual scatter reflects variations in halo concentration, ICL formation efficiency, and the detailed assembly history of the system.

Two additional criteria regulate this process. First, if the transition radius $R_{\rm{trans}}$ lies within the bulge radius, all SH stars within it are transferred to the bulge. Second, for stability reasons, the SH mass cannot exceed the stellar mass of the BCG (bulge+disk) \footnote{Bulge and disk formation and evolution are implemented in FEGA as done in \citep{guo2011}}. Any excess mass in the SH is redistributed to the BCG’s disk. These stability criteria imply that the SH mass is not a fixed fraction of the ICL mass, even at fixed halo mass or concentration, and introduce physically motivated scatter in all SH-related scaling relations.

\begin{figure}
\centering
\includegraphics[width=0.48\textwidth]{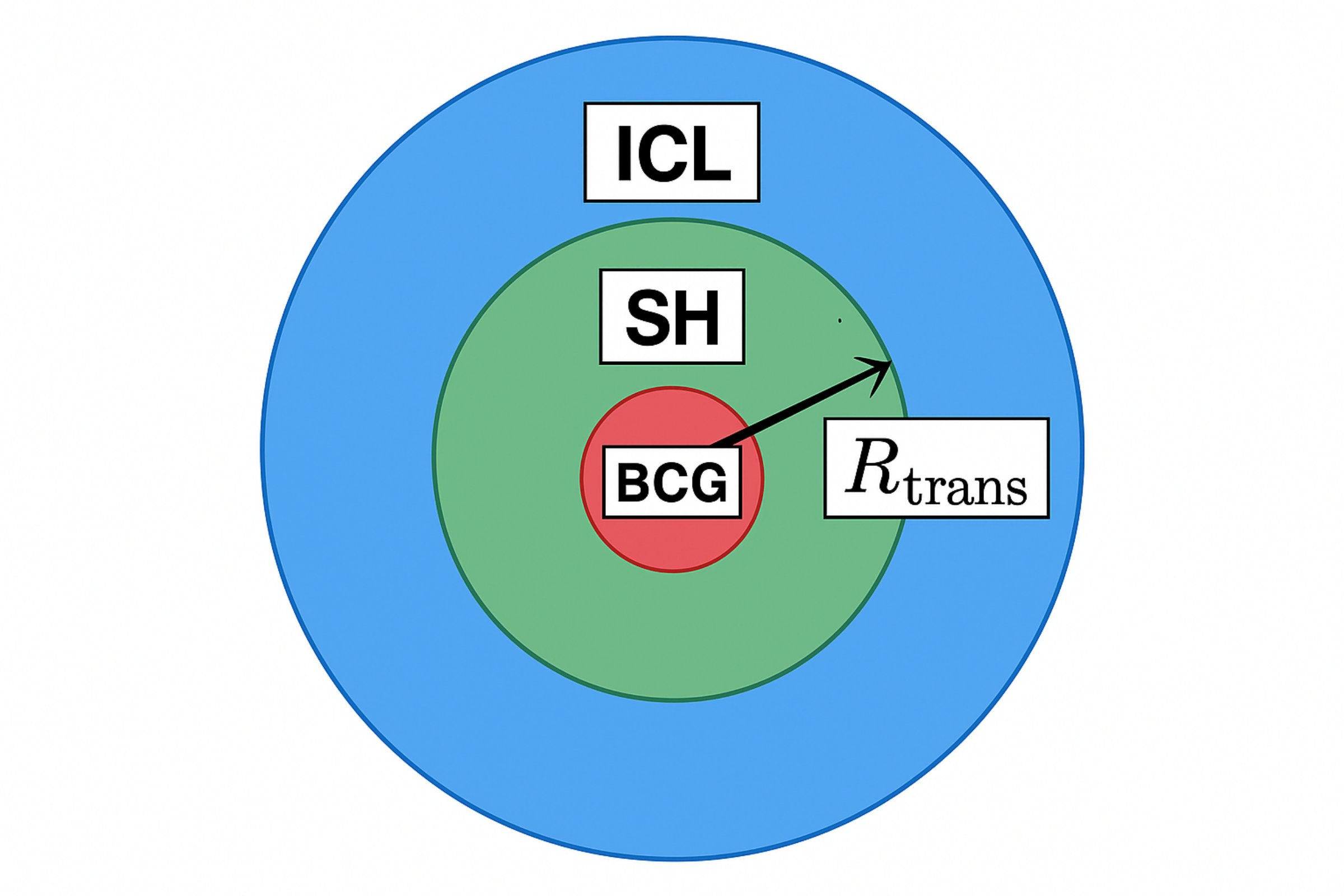}
\caption{Schematic representation of central galaxies and their surrounding stellar halos and intracluster stars. The stellar halo is defined as the stellar component within the transition radius $R_{\rm{trans}}$, marking the intermediate region between the galaxy and the intracluster light.}
\label{fig:scheme}
\end{figure}

As highlighted above, the SH stars may or may not be bound to the central galaxy. In the first case, the SH can be interpreted as a third component of the BCG+SH+ICL system, while in the second case—or in the mixed case with both bound and unbound stars—it represents a transition region between the galaxy and the surrounding ICL. Following \citet{contini2024d}, where SHs were first implemented, in this work we adopt the latter interpretation. Thus, the system is modelled as a central galaxy, consisting of bulge and disk, surrounded by a SH, all embedded in a more extended diffuse light component (see Figure~\ref{fig:scheme} for a schematic illustration). We therefore emphasize that the interpretation of SHs as transition regions is a direct consequence of the adopted definition, while the quantitative trends in their mass, structure, and stellar populations explored in Section~\ref{sec:results} reflect genuine physical dependencies on halo concentration, ICL assembly, and satellite accretion.

Although the definition of the transition radius is somewhat arbitrary, in \citet{contini2024d} we showed that the typical values of $R_{\rm{trans}}$ for halos of given mass are consistent with both observations and theoretical predictions, such as the relation found by \citet{proctor2024}. It is also worth discussing an important caveat. The observational definition of the SH is not unique. Historically, the integrated light profiles of the Milky Way and other disk galaxies were successfully fit by a de Vaucouleurs component dominating both the central regions and the outskirts, which led to the interpretation of the SH as a photometric extension of the bulge (e.g., \citealt{deVaucouleurs1959}). A purely morphological classification does not capture the physical distinctiveness of the halo, whose stars typically exhibit low metallicities, dispersion-dominated kinematics, and signatures of accretion and minor merging (e.g., \citealt{carollo2007,carollo2010,helmi2008,deason2013,belokurov2020}). This is particularly clear in the Milky Way, where Gaia has revealed that a significant fraction of the inner halo originates from a major accretion event (Gaia-Enceladus; \citealt{helmi2018,belokurov2018}), and where chemical and kinematic surveys identify multiple substructures and a dual-halo configuration. In this work we therefore adopt a physical definition of the SH, based on dynamical selection criteria rather than on a decomposition of the light profile alone. This approach reflects modern observational evidence that the halo traces the hierarchical assembly history of the host galaxy, and is distinct from bulge and disk components both in origin and in stellar population properties (e.g., \citealt{gallart2019,naidu2020}).

\subsection{Stellar populations: colours and metallicities}
\label{subsec:stellar_pops}

The model tracks the age and metallicity of the stellar populations associated with each stellar component (BCG, SH, and ICL) by following their star-formation and chemical-enrichment histories self-consistently within the semi-analytic framework. Broad-band colours are computed by applying stellar population synthesis models to the predicted stellar mass, age, and metallicity of each component, assuming a Chabrier initial mass function.

In this approach, colours and metallicities are not free parameters tuned to reproduce the observations, but are direct outputs of the model, inherited from the formation histories of the different stellar components. In particular, the SH and the ICL are assumed to evolve passively after their formation, with no \emph{in situ} star formation, consistently with their predominantly accreted origin.

Mean metallicities are derived using the same mass-weighting scheme adopted for the computation of colours, ensuring a coherent mapping between the semi-analytic outputs and the
observable quantities used in the comparison with data. While stellar populations represent one of the most challenging aspects to model within a semi-analytic framework, this methodology allows us to perform a physically motivated and internally consistent comparison between model predictions and observations.

In the remainder of this section, we briefly describe the merger trees (Section~\ref{sec:trees}) and the observational data set (Section~\ref{sec:data}) before moving on to the analysis and discussion of the main results.

\subsection{Merger trees}\label{sec:trees}

\begin{table}
\begin{center}
  \caption{Name of the simulation (first column), box size (second column), resolution (third column) and thresholds in selecting dark matter halos (fourth column).}
\label{tab:simulations}
\begin{tabular}{lccccc}
\hline
\hline
&Name &L [Mpc/h] &Resol. [$M_{\odot}/\rm{h}$] &Thresh. [$M_{\odot}/\rm{h}$]\\
\hline
&YS300  &300  & $2.2\cdot10^9$    & $>10^{14}$ \\
&YS200  &200  & $3.26\cdot10^8$   & $>10^{13}$ \\
&YS50HR &50   & $10^7$            & $>10^{12}$ \\
\hline
\hline
 \end{tabular}
\end{center}
\end{table}

In order to construct galaxy catalogs, semianalytic models require merger trees extracted from dark matter–only numerical simulations. For this purpose, we employ a set of three cosmological simulations carried out with the latest version of the {\small GADGET} code, namely {\small GADGET-4} \citep{springel2021}. These simulations span different volumes and mass resolutions, as summarized in Table~\ref{tab:simulations}, and the corresponding merger trees have been used to generate the inputs for {\small FEGA25}.

All runs adopt the same comoving gravitational softening length of $3\,{\rm kpc}/h$ \footnote{Within the semi-analytic framework adopted here, the gravitational softening length does not directly affect the predicted galaxy structural properties, which are computed analytically rather than resolved numerically. To further minimize resolution-driven effects, each simulation is therefore used only within the halo-mass regime where it provides a reliable sampling of the underlying population, and no direct comparison of the same systems across different resolutions is performed in this work.}, and follow the Planck 2018 cosmology \citep{planck2020}: $\Omega_m = 0.31$ for the matter density parameter, $\Omega_{\Lambda} = 0.69$ for the cosmological constant, $n_s = 0.97$ for the primordial spectral index, $\sigma_8 = 0.81$ for the power spectrum normalization, and $h = 0.68$ for the dimensionless Hubble parameter. The simulations were evolved from an initial redshift of $z=63$ down to the present day, with data stored in 100 snapshots uniformly distributed in cosmic time between $z=20$ and $z=0$.

We note that numerical resolution can play an important role when the same halos are simulated at different resolutions, particularly for quantities related to satellite survival,
stripping, and mass loss (e.g. \citealt{contini2014}). However, this effect has been explicitly tested in previous work, where we verified that, in the halo-mass ranges where different
simulations overlap, the predicted properties of diffuse stellar components are consistent within the intrinsic scatter. Nevertheless, in Appendix \ref{sec:app_resolution} we provide some additional plots, separating the data coming from the different runs.

Because of the different resolutions of the simulations, we impose distinct mass thresholds when selecting dark matter main halos (focusing exclusively on centrals), as listed in Table~\ref{tab:simulations}. The largest volume run, YS300, provides a statistically significant sample of massive groups and clusters, while the highest-resolution box enables us to probe Milky Way–sized halos. This combined catalog forms the basis for the analysis presented in Section~\ref{sec:results}.

\subsection{Observed data}\label{sec:data}

To compare the predictions of our model with observational data, we make use of the survey VEGAS \citep{capaccioli2015,iodice2021} and the Fornax Deep Survey (FDS; \citealt{iodice2016}). These surveys do not provide metallicity measurements. Therefore, for the same targets, we complement them with data from the Fornax3D project (F3D; \citealt{sarzi2018}) and the M3G survey \citep{krajnovic2018}. Throughout the rest of the paper, we refer to this combined dataset simply as VEGAS, which includes both colours and metallicity information.

VEGAS is a wide-field deep imaging program conducted with the VLT Survey Telescope (VST) at ESO’s Paranal Observatory. Its primary aim is to explore the faint outskirts of nearby ETGs down to very low surface brightness levels, typically reaching $\mu_g \sim 28.5$–$29$ mag arcsec$^{-2}$ (\citealt{iodice2021}). By combining multi-band wide-field and deep optical imaging (\emph{u, g, r, i}), VEGAS allows for a detailed study of the structural properties and stellar populations in the outer regions of galaxies, unveiling faint SHs, extended envelopes, and substructures such as shells, streams, and tidal debris, which are relics of past accretion and merging events \citep{mihos2017}.

In addition, VEGAS plays a crucial role in the investigation of the ICL. This component provides direct constraints on the assembly history of massive structures and the efficiency of tidal stripping and galaxy interactions (\citealt{contini2014}; \citealt{montes2018}; \citealt{spavone2020}). By building a statistical sample of ETGs across diverse environments, from isolated systems to dense clusters, VEGAS delivers a key observational benchmark for hierarchical galaxy formation models and for the study of the stellar mass assembly at large radii (\citealt{cooper2010}; \citealt{pillepich2018}). For further details and references about the survey, we refer the reader to \citet{contini2024d}.

The Fornax Deep Survey (FDS) is a deep optical imaging program carried out with the VST as part of the VEGAS collaboration. Its main goal is to map the entire Fornax cluster down to unprecedented surface-brightness levels of about $\mu_g \sim 29$ mag arcsec$^{-2}$, thus providing the first homogeneous and contiguous photometric coverage of both the cluster core and its periphery (\citealt{iodice2016}). With a wide field of view of one square degree per pointing and multi-band coverage (\emph{u, g, r, i}), FDS enables a systematic exploration of the galaxy population of Fornax, ranging from early- and late-type systems to ultra-diffuse galaxies and low-surface-brightness dwarfs (\citealt{venhola2017}; \citealt{venhola2018}).

A key strength of FDS is its ability to trace the ICL and the extended stellar envelopes of massive galaxies, which represent direct evidence of the cluster’s ongoing assembly and provide insights into tidal stripping, galaxy harassment, and merging processes (\citealt{iodice2017}; \citealt{spavone2017}). Moreover, by resolving faint structural and colour gradients across galaxies, the survey constrains the stellar populations and metallicity distributions, thereby shedding light on environmental transformation processes and their role in shaping galaxy evolution within dense environments.

FDS stands as one of the deepest wide-area optical surveys of a nearby cluster, complementing similar campaigns in Virgo and Coma. By simultaneously resolving the brightest galaxies and the faintest dwarf systems, it delivers a comprehensive view of galaxy evolution in clusters and offers crucial observational constraints for cosmological models of structure formation (\citealt{drinkwater2001}; \citealt{capaccioli2015}). For further details about FDS we refer the reader to \citet{iodice2016}.

We note that, while VEGAS primarily targets bright central galaxies, the FDS sample includes a significant fraction of satellite systems. For this reason, FDS data are not used here as a direct counterpart of the simulated sample, which includes only central galaxies. Instead, FDS provides a complementary reference for the properties of diffuse outer stellar components in a dense cluster environment. The comparison with model predictions is therefore statistical in nature and is not intended as a one-to-one correspondence between individual observed and simulated systems. This comparison therefore provides a consistency check on the typical scales and trends of diffuse stellar components, rather than a direct test of individual galaxy properties. In the following analysis, we remind the reader that we combine VEGAS-FDS-F3D-M3G data and, for simplicity, refer to the entire dataset as VEGAS in the discussion, but we separate them in the analysis where VEGAS data alone will be indicated as VG.

For the full sample of galaxies, we consider the \emph{g-r} and \emph{r-i} colours. Moreover, we define two transition radii, $R_1$ and $R_2$, as the radii at which the second and third components of the 1D surface-brightness profile fits begin to dominate over the first and second components, respectively. In this work, when comparing with the model-defined transition radius, we adopt $R_2$, which marks the transition between the accreted bound and the diffuse unbound stellar components. This choice is motivated by the fact that $R_2$ represents the most appropriate observational analogue of the SH-ICL boundary probed by the semi-analytic model.  Not having the \emph{g-r} or \emph{r-i} colours available for all targets (for around 40\% of them), nor the second transition radius $R_2$ (for around 20\% of them), we adopted an empirical calibration strategy based on a subset of galaxies for which both radii and multi--colour photometry were available. Specifically, we proceeded as follows:

\begin{enumerate}
    \item Training sample.

    From the galaxies with both $R_1$ and $R_2$ measured, we converted angular radii from arcminutes into arcseconds (and subsequently into kiloparsecs at the distance of the Fornax cluster, assuming $1'' \simeq 0.097$ kpc at 20 Mpc). Missing photometric values in the \emph{g-r} and \emph{r-i} colours were treated via a bootstrap imputation scheme \citep{efron1979}:
    \begin{itemize}
        \item If one colour was available, the other was inferred from a linear regression relation calibrated on galaxies with both colours, adding scatter consistent with the observed residuals.
        \item If both colours were missing, we imputed values from the bootstrap distribution of the sample mean.
    \end{itemize}
    This procedure, repeated for 5000 bootstrap resamplings, provided a median estimate together with 16th--84th percentile intervals for each imputed value.
\vspace{0.2cm}
    \item Empirical $R_1 \to R_2$ mapping.

    We modeled the correlation between the two transition radii in logarithmic space,
    \[
    \log_{10} R_2 = a + b \,\log_{10} R_1 \, ,
    \]
    and calibrated the parameters $(a,b)$ from the training sample \footnote{The adoption of a power-law relation between $R_1$ and $R_2$ is motivated by the fact that both quantities trace
characteristic transition scales in multi-component surface- brightness profiles and are therefore expected to scale approximately proportionally in logarithmic space.}. A bootstrap approach (5000 realizations) was used to propagate both coefficient uncertainties and intrinsic scatter.
\vspace{0.2cm}
    \item Application to target galaxies.

    For observed galaxies with only $R_1$, we predicted $R_2$ using the calibrated relation. Each prediction was expressed as a median with an associated 16th--84th percentile range, thus quantifying the uncertainty of the extrapolation. Alongside the structural radii, we retained the outer stellar metallicity to allow further analysis of correlations between halo properties and stellar populations.
\end{enumerate}

\section{Results and discussion}\label{sec:results}

In the following, we present our analysis, which is organized into two main parts. In the first part, we examine the key scaling relations between the SH mass and the transition radius $R_{\rm trans}$ defined above, as well as with the masses of the BCG and the ICL. In this context, we also assess the role of halo concentration and the efficiency of ICL formation, two fundamental factors in shaping SHs. In the second part, we focus on the colours (\emph{g–r} and \emph{r–i}) and metallicity of the three components (BCGs, SHs, and ICL). In both cases, we trace their redshift evolution from $z=2$ to the present day.

\subsection{Scaling relations}\label{sec:relations}

\begin{figure*}[t!]
\begin{center}
\begin{tabular}{cc}
\includegraphics[width=0.47\textwidth]{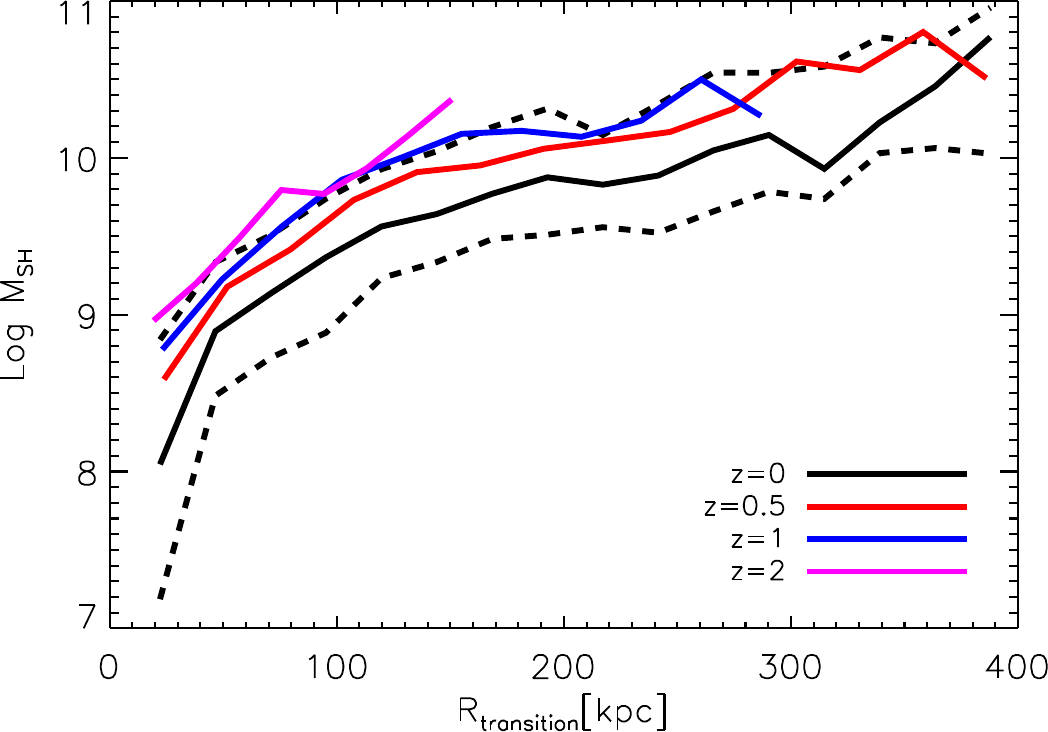} &
\includegraphics[width=0.475\textwidth]{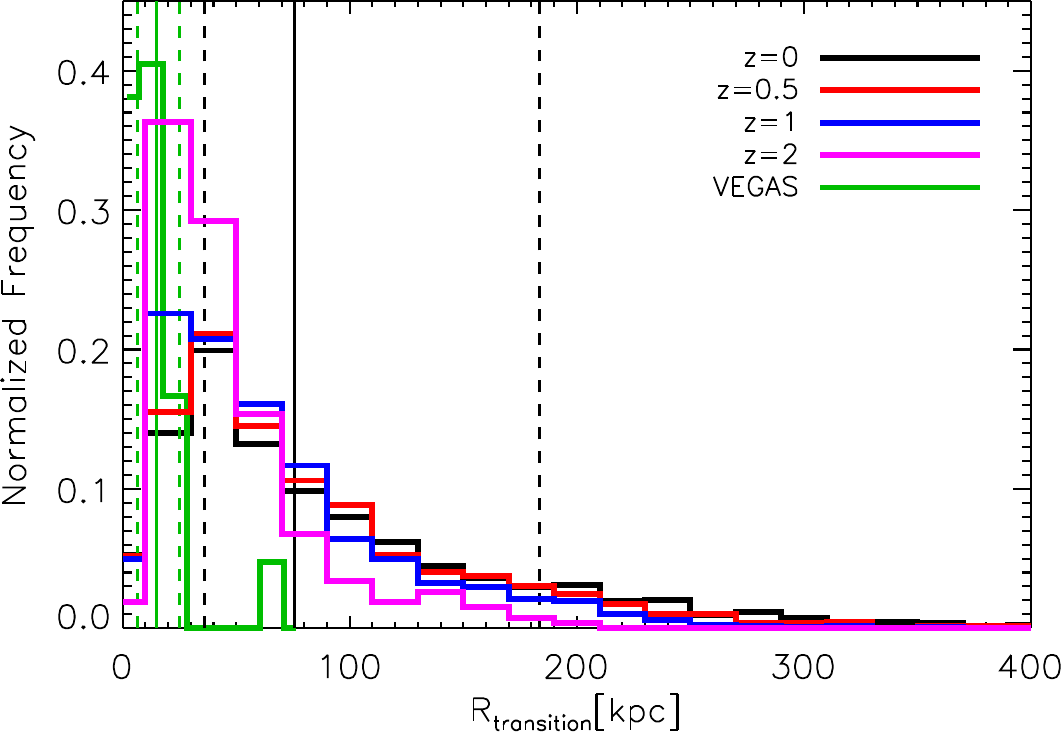} \\
\end{tabular}
\caption{Left panel: stellar halo mass as a function of the transition radius $R_{\rm{trans}}$, at different redshifts as indicated in the legend. The two black dashed lines represent the 16th and 84th percentiles of the distribution at $z=0$. At fixed $R_{\rm{trans}}$, the stellar halo mass increases slightly with redshift, since dark matter halos with similar concentrations are more evolved at earlier epochs than at later times. Right panel: Distribution of transition radii at different redshifts (different colours) and as inferred from observations (VEGAS, green line). The model distributions extend to systematically larger values of $R_{\rm trans}$ than the observational one, while the VEGAS distribution is more narrowly peaked at smaller radii (see the vertical lines indicating medians, 16th and 84th percentiles of the distributions for the model predictions at $z=0$ (black) and VEGAS (green)). This offset reflects the different definitions adopted in the SAM and in observations, as well as differences in sample selection. The comparison should therefore be interpreted in a statistical sense, focusing on typical scales and trends rather than on a one-to-one correspondence. Across all redshifts, the majority of model transition radii lie within $\sim 100$ kpc, with a tail extending to larger values in the most massive systems.}
\label{fig:rtrans}
\end{center}
\end{figure*}

The transition radius $R_{\rm trans}$ is a fundamental quantity in defining SHs, particularly their mass, which also depends on the total stellar mass contained in the ICL. Importantly, $R_{\rm trans}$ is strongly determined by the ICL concentration, which in turn is tightly connected to the halo concentration. In the left panel of Figure~\ref{fig:rtrans}, we show the SH mass as a function of $R_{\rm trans}$ at different redshifts (colours as indicated in the legend). The dashed lines represent the 16th and 84th percentiles of the distribution at $z=0$. The figure reveals that this relation depends only weakly on redshift: only the $z=2$ curve falls outside the upper percentile envelope of the $z=0$ distribution.

Interestingly, at fixed $R_{\rm trans}$, the SH mass increases with redshift. A constant transition radius implies the same ICL concentration, and a halo concentration that is a factor $\gamma$ (between 1 and 3; see Section~\ref{sec:methods}) lower. Given the halo mass–concentration relation (e.g., \citealt{prada2012,correa2015,child2018}), this suggests that BCG host halos at different epochs may exhibit similar concentrations within the intrinsic scatter of the relation. At higher redshift, such concentrations correspond to dynamically more evolved halos, which likely produced a substantial amount of ICL, part of which is converted into significant SH mass. We verified this scenario by selecting samples of halos at $z=2$ and comparing them with their $z=0$ counterparts at fixed $R_{\rm trans}$, finding that this trend is indeed statistically robust.

Another point worth stressing is that $R_{\rm trans}$ can extend to very large scales: at $z=0$ it can reach $\sim 400$ kpc in the most massive halos, while $\sim 300$ kpc at $z=1$, and at $z=2$ it still extends up to $\sim 150$ kpc. The systematic decrease of $R_{\rm trans}$ with redshift is a natural outcome of the hierarchical growth of structures: halos at higher redshifts are, on average, less massive and dynamically younger. At least at $z=0$, where a comparison is possible, our results are in good agreement with those of \citet{proctor2024}, who analyzed the EAGLE (\citealt{schaye2015,crain2015}) and C-EAGLE (\citealt{bahe2017,barnes2017}) simulations using an independent definition of $R_{\rm trans}$ \citep{contini2024d}.

The right panel of Fig.~2 shows the distribution of transition radii, $R_{\rm trans}$, predicted by the model at different redshifts and inferred from the VEGAS observations. While the observational distribution is narrowly peaked at relatively small radii, the model predicts a broader distribution extending to systematically larger values, particularly at low redshift. This offset indicates that observationally defined transition radii, derived from multi-component photometric decompositions through minimization of the rms scatter of the fits (e.g. \citealt{seigar2007}), and shown to correlate with changes in surface-brightness, ellipticity, position angle and colour profiles \citep{spavone2017,spavone2021}, do not necessarily trace the same physical boundary as the model-defined $R_{\rm trans}$, which separates stars bound to the central galaxy from the diffuse intracluster component. \footnote{For the observational comparison shown in Figure~\ref{fig:rtrans}, we use the $R_2$ values derived from VEGAS, corresponding to the transition between the accreted bound and the diffuse unbound stellar components.}

Part of the difference may reflect sample selection effects, as the observational data probe a restricted range of massive systems in the local Universe, whereas the model includes central galaxies spanning a wider range of stellar and halo masses. In addition, the heterogeneous distances of the VEGAS targets introduce an additional source of scatter in the conversion from angular to physical radii. For these reasons, the comparison between observations and model predictions should be interpreted in a statistical and qualitative sense, focusing on typical scales and trends rather than on a direct, one-to-one correspondence.

\begin{figure*}[t!]
\begin{center}
\begin{tabular}{cc}
\includegraphics[width=0.47\textwidth]{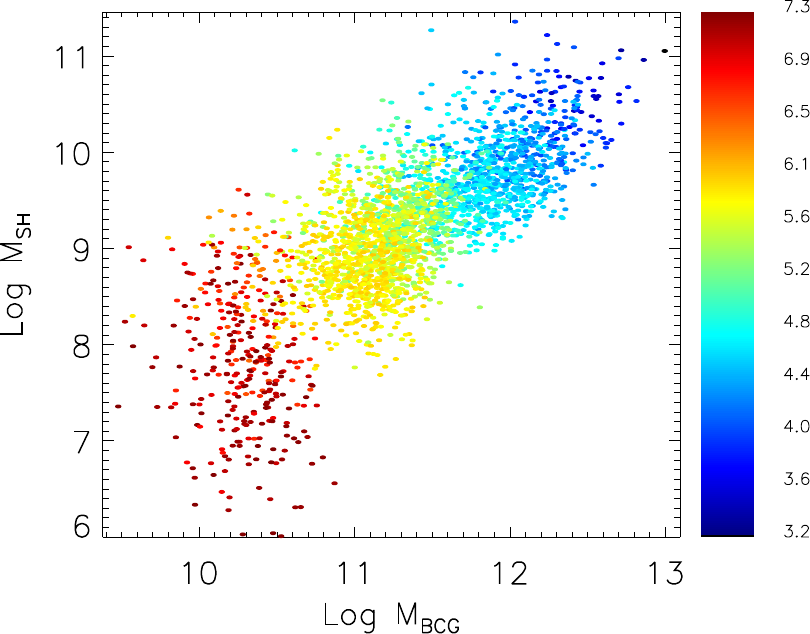} &
\includegraphics[width=0.47\textwidth]{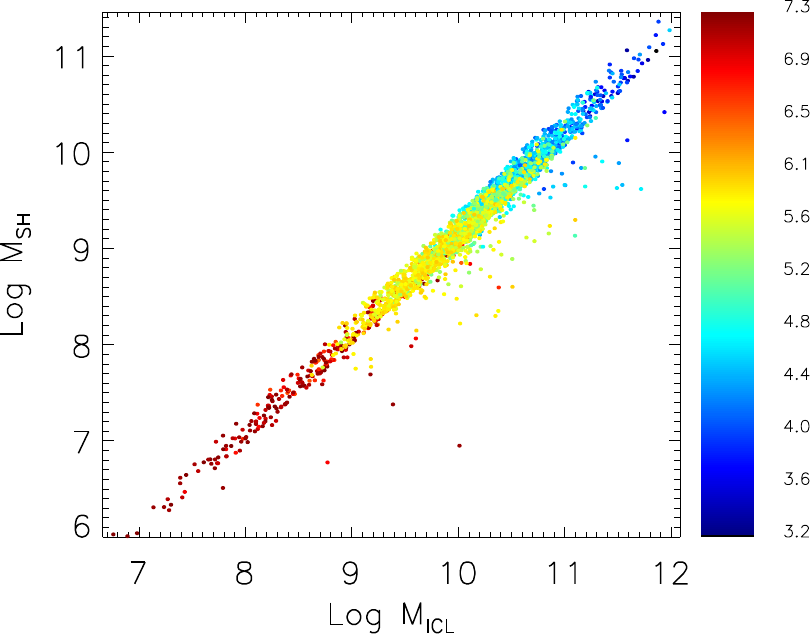} \\
\end{tabular}
\caption{Scaling relations between the mass of stellar halos and that of the associated central galaxy (left panel) and intracluster light (right panel). The colour bar encodes the dark matter halo concentration, with redder colours corresponding to higher concentrations. The stellar halo mass correlates well with both the BCG and the ICL mass, with a notably smaller scatter in the ICL case. In both relations, halo concentration plays a key role: stellar halos tend to be more massive in less concentrated dark matter halos, which are also the most massive systems.}
\label{fig:scale1}
\end{center}
\end{figure*}

\begin{figure*}[t!]
\begin{center}
\begin{tabular}{cc}
\includegraphics[width=0.47\textwidth]{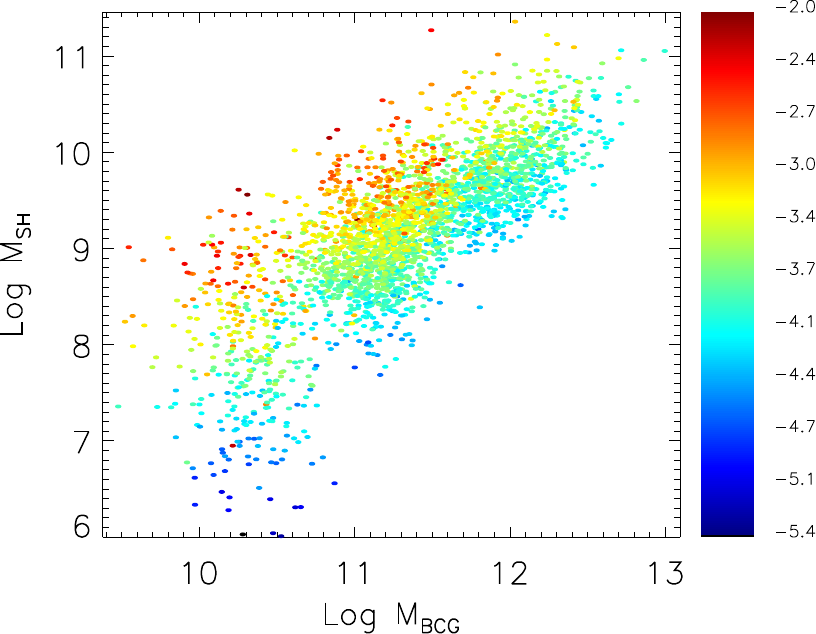} &
\includegraphics[width=0.47\textwidth]{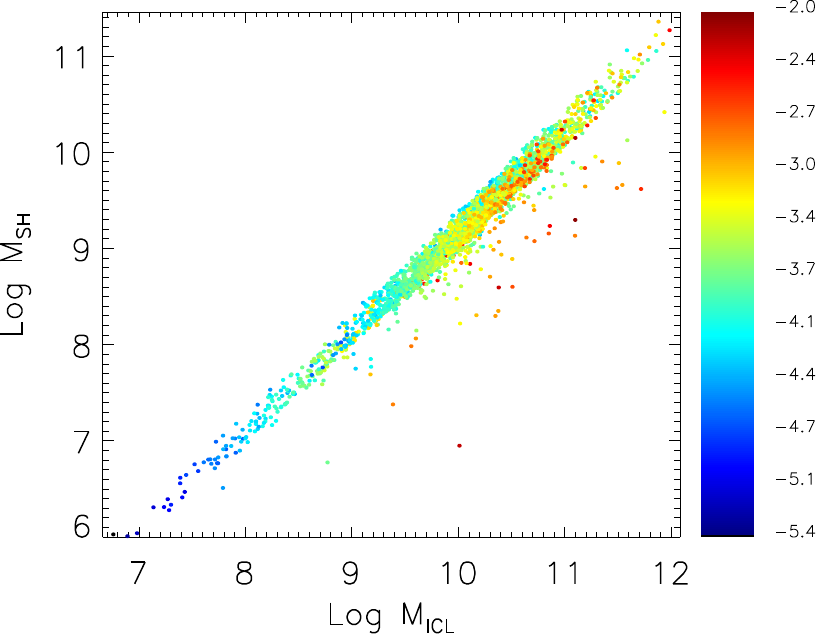} \\
\end{tabular}
\caption{Similar to Figure~\ref{fig:scale1}, the plots show the same scaling relations, but with the colour bar now indicating the logarithm of the efficiency of ICL production in dark matter halos. The efficiency is defined as the ratio between the ICL mass and the halo mass. A clear trend emerges in both panels: stellar halos tend to be more massive when the efficiency of ICL production is higher. In the left panel, this appears at fixed BCG mass, where redder colours correspond to more massive stellar halos, while in the right panel the trend follows the main relation itself.}
\label{fig:scale2}
\end{center}
\end{figure*}

In Figure~\ref{fig:scale1} we present the scaling relations between the SH and the BCG mass (left panel), and between the SH and the ICL mass (right panel), at $z=0$. In both cases, circles are colour coded according to the halo concentration. We find that the SH mass correlates with both the BCG and the ICL mass, although the latter relation exhibits a significantly smaller scatter. This is not surprising, given that SHs form directly from stars belonging to the ICL, or because the SH is a tail of the ICL instead of a separate component. An important trend visible in both panels is that the SH mass tends to increase in less concentrated halos, which are also typically the most massive ones. At first sight, this might appear in tension with the results of \citet{contini2023}, who found higher ICL fractions in more concentrated halos. However, it should be stressed that our plots show mass-to-mass relations, not fraction-to-mass relations. Moreover, halo concentration is directly linked to the transition radius $R_{\rm trans}$. Less concentrated halos (and their associated ICL) are characterized by larger values of $R_{\rm trans}$ (see Equation~\ref{eqn:rtrans}), which in turn implies a larger SH mass, depending on the amount of ICL already in place. The key message is therefore that halo concentration, through its connection with the ICL concentration, plays a crucial role in shaping the formation of SHs.

Another striking feature is that both scaling relations appear nearly linear by eye. Focusing on the SH–BCG mass relation, we performed a simple linear fit in logarithmic space. The best-fit slope is 0.99, with a relatively small intercept of $-2.07$, confirming that the relation is essentially linear. This allows us to reformulate the fit in terms of the mass fraction $f_{\rm SH-BCG}$ on the $y$-axis, since the $f_{\rm SH-BCG}$–BCG mass relation is nearly constant. Within the range of $\log M_{\rm BCG}=[9.5,13]$, we obtain the following values: mean $f_{\rm SH-BCG}$ = [$0.0074$, $0.0069$], mean$+1\sigma$ = [$0.0158$, $0.0174$], mean$-1\sigma$ = [$0.0034$, $0.0028$], mean$+2\sigma$ = [$0.0347$, $0.0427$], and mean$-2\sigma$ = [$0.0016$, $0.0011$]. These predictions compare very well with the observed data from \citet{merritt2016} (their Figure 5, left panel) and \citet{gilhuly2022} (their Figure 7). Both works are based on the Dragonfly Survey, with Gilhuly et al. extending the analysis to lower-mass galaxies than Merritt et al.. Specifically, our $\pm 1\sigma$ region encompasses the bulk of the Merritt et al. data, with the exception of three outliers not even included within $\pm 2\sigma$. Similarly, for the Gilhuly et al. sample, our $\pm 1\sigma$ region accounts for most of their galaxies, while $\pm 2\sigma$ essentially covers the full set. This agreement provides strong support for the robustness of our SH implementation in {\small FEGA25}.

While halo concentration is a fundamental driver of ICL and SH formation, another quantity that may play a role is the efficiency of ICL production. Intuitively, the larger the amount of ICL, the more likely it is to form massive SHs. We explore this aspect in Figure~\ref{fig:scale2}, which shows the same two relations as in Figure~\ref{fig:scale1}, but this time with circles colour coded by the ICL formation efficiency. We define the efficiency as the ratio between the total ICL mass assembled up to a given epoch (here $z=0$) and the total halo mass. As shown in \citet{contini2024}, the efficiency of ICL formation is broadly halo-mass independent, although this does not necessarily preclude a role in SH growth.

Both panels in Figure~\ref{fig:scale2} highlight a clear trend: SHs are more massive in systems with higher ICL formation efficiencies. This is most evident in the left panel, where at fixed BCG mass (and thus at approximately fixed halo mass within a range), redder circles indicate larger efficiencies. In the SH–ICL mass plane, the trend is somewhat less pronounced but still visible, with low-mass SHs corresponding to bluer (lower-efficiency) points, and high-mass SHs associated with redder ones. Taken together, halo concentration and ICL formation efficiency both emerge as key factors in shaping SH properties. We also emphasize that repeating this analysis at higher redshifts leads to no significant differences: the trends already hold at earlier times, with the expected variations in scatter due to number statistics.

Having explored these scaling relations, we now turn to more challenging properties to model in a semianalytic framework, namely colours and metallicities.

\subsection{Colors and metallicity}\label{sec:colmet}

\begin{figure*}[t!]
\begin{center}
\begin{tabular}{cc}
\includegraphics[width=0.47\textwidth]{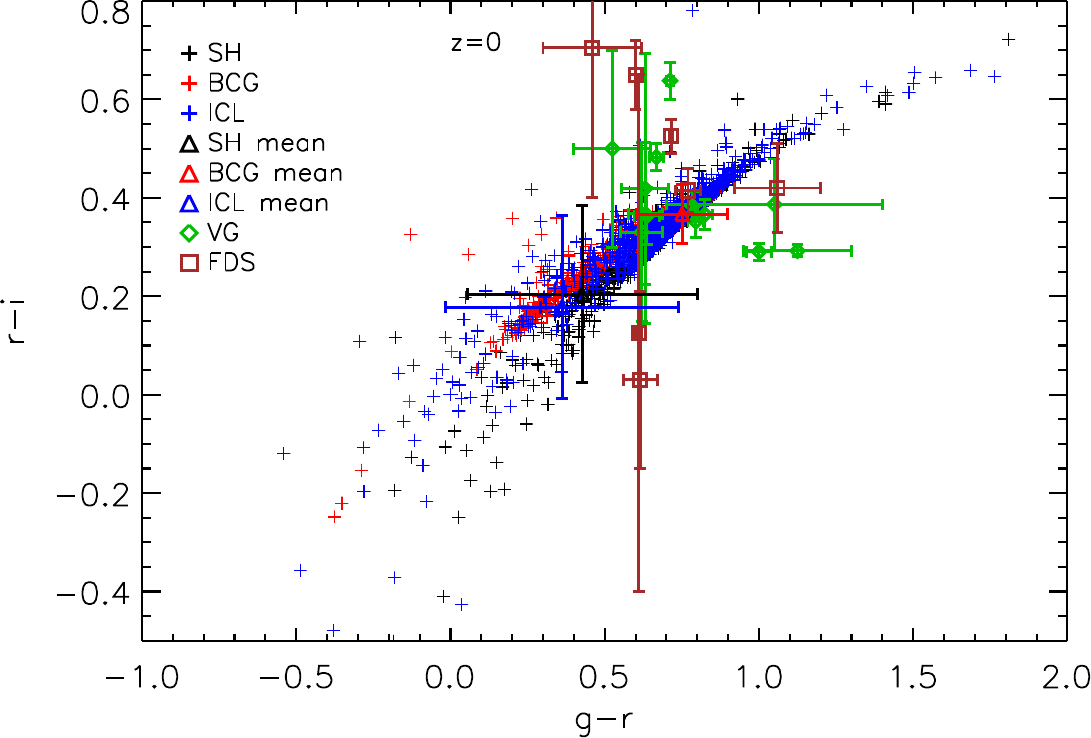} &
\includegraphics[width=0.47\textwidth]{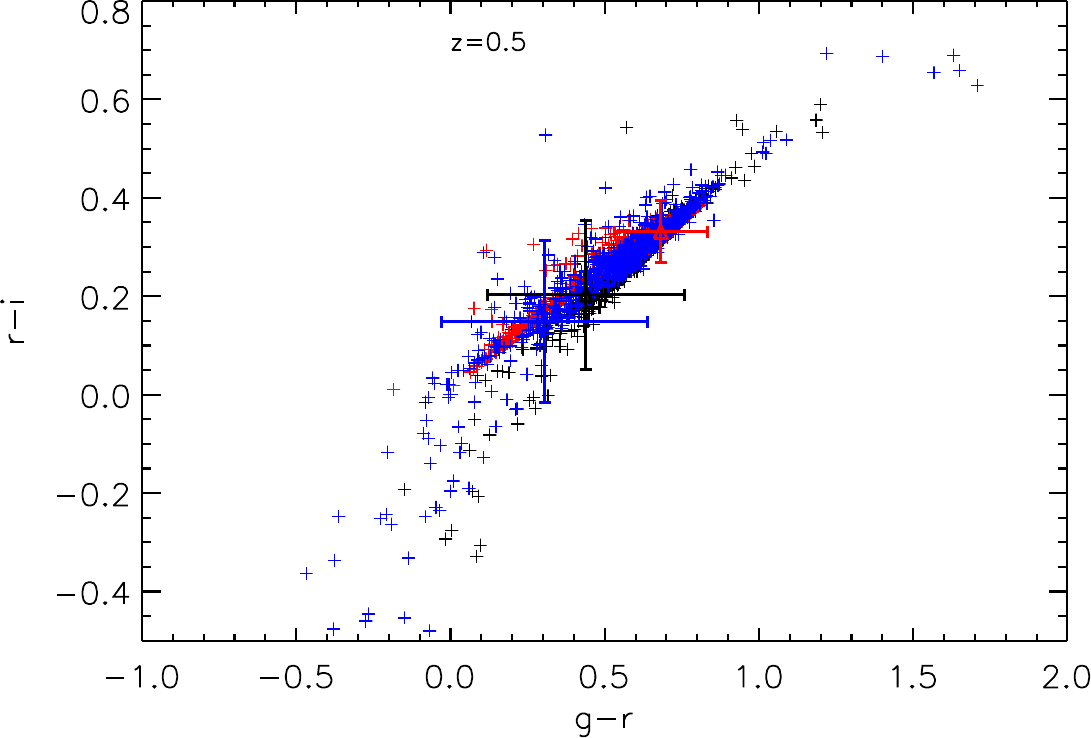} \\
\includegraphics[width=0.47\textwidth]{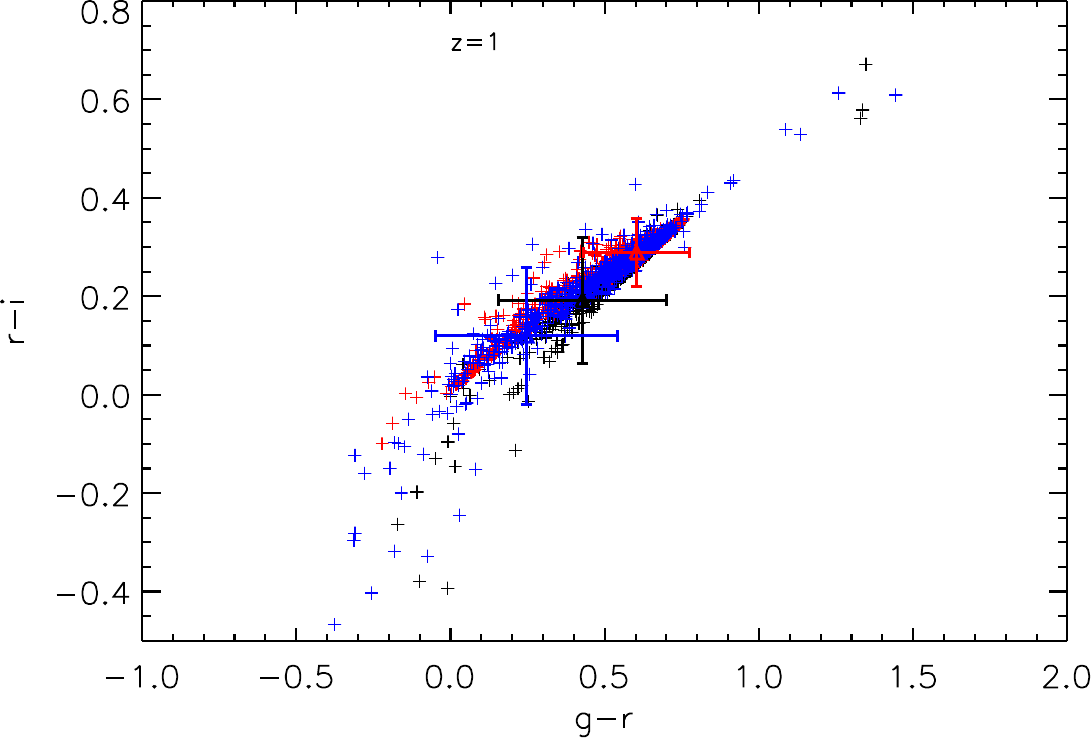} &
\includegraphics[width=0.47\textwidth]{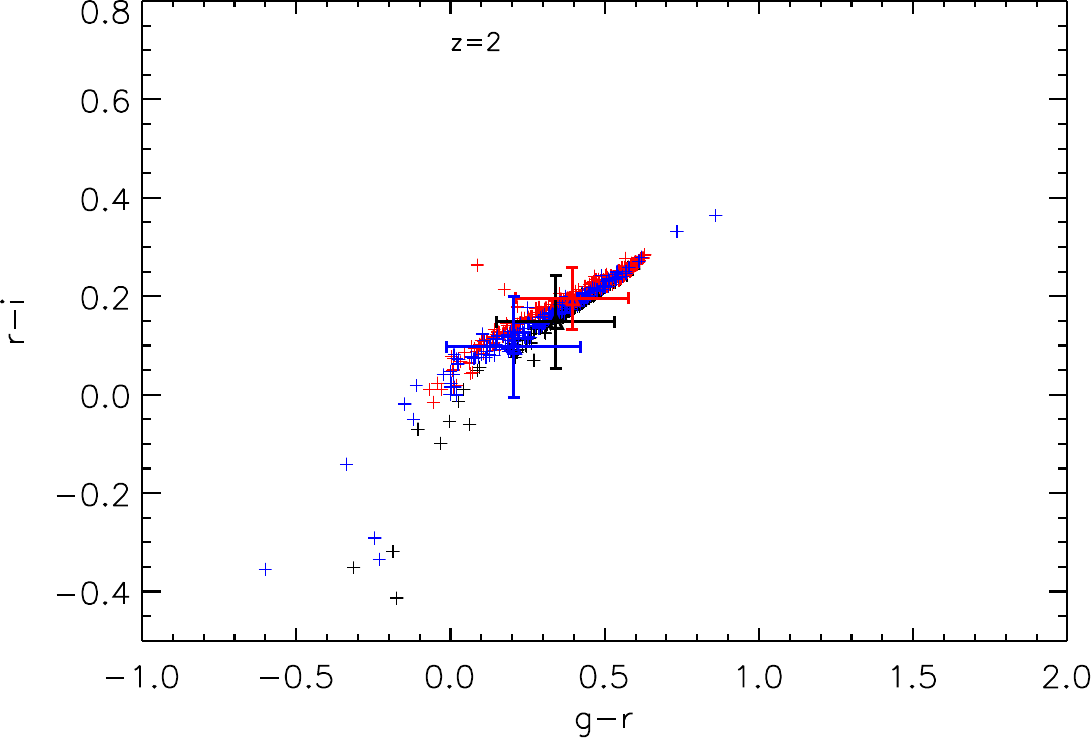} \\
\end{tabular}
\caption{Relation between \emph{g-r} and \emph{r-i} colours for the BCGs (red), stellar halos (black), ICL (blue) at different redshifts (separate panels), and as observed in VEGAS (green diamonds for VG and brown squares for FDS, at $z=0$). Overall, the three components exhibit comparable colours, largely independent of redshift, and all progressively redden (shifting towards the right side of the diagrams) as the redshift approaches the present epoch.}
\label{fig:colcol}
\end{center}
\end{figure*}

\begin{figure*}[t!]
\begin{center}
\begin{tabular}{cc}
\includegraphics[width=0.47\textwidth]{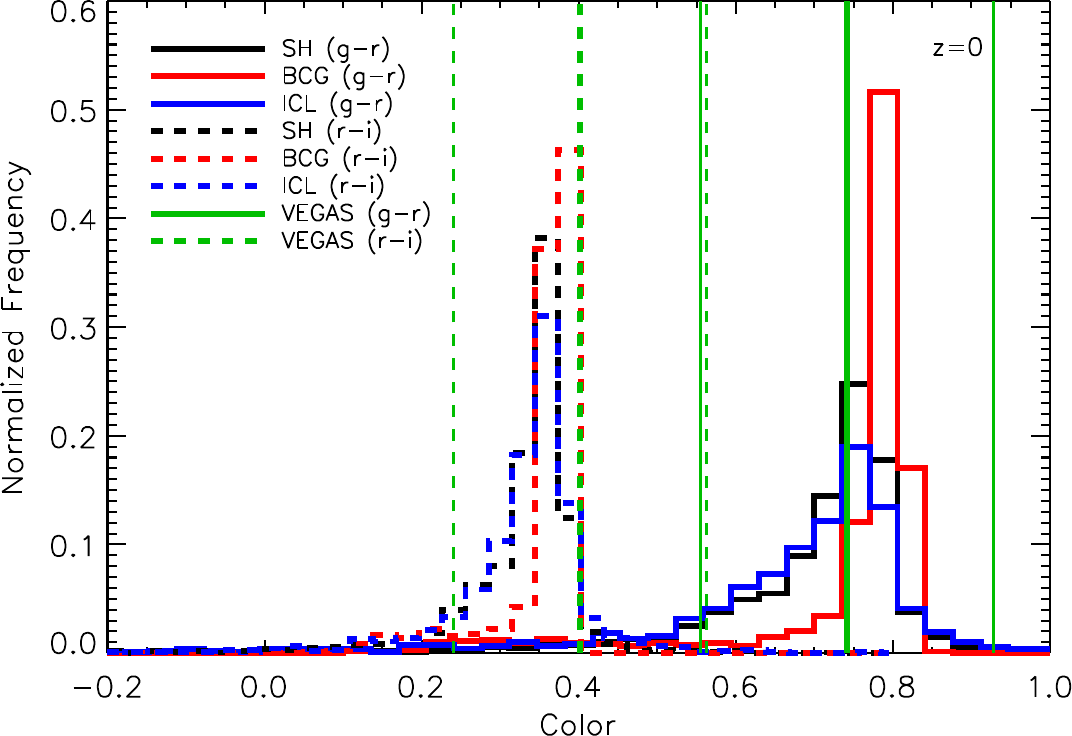} &
\includegraphics[width=0.47\textwidth]{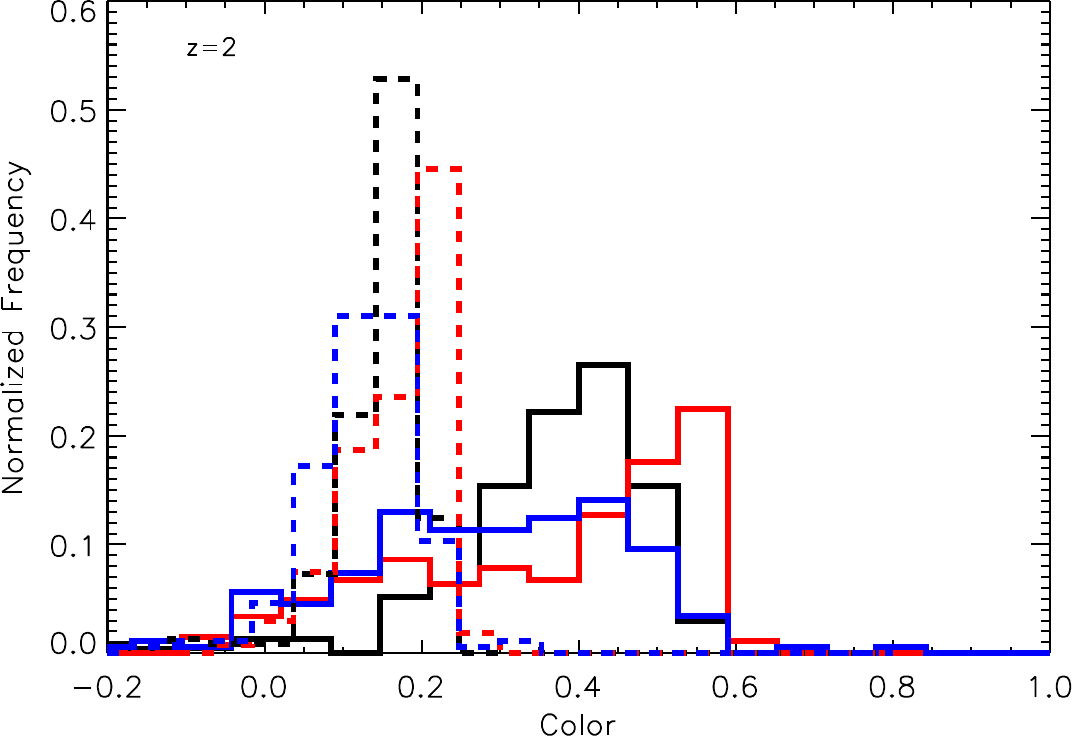} \\
\end{tabular}
\caption{For the same colours shown in the previous figure, the plots illustrate their distributions for each component at $z=0$ (left panel) and $z=2$ (right panel). As discussed above, both colours become redder with time, but the distributions reveal notable differences. At $z=0$, the BCGs display narrower distributions with a more pronounced peak, while at $z=2$ their distributions appear quite different. Overall, stellar halos and the ICL exhibit similar distributions at $z=0$, whereas at $z=2$ they are statistically distinct. Importantly, BCGs are consistently redder than the other two components, independent of redshift. The green thick lines (solid and dashed) represent the mean colours observed in VEGAS (VG+FDS), while the thinner ones indicated the $\pm 1\sigma$ distribution. The predicted \emph{g-r} and \emph{r-i} of the SHs at $z=0$ are very close to the observed ones.}
\label{fig:coldistr}
\end{center}
\end{figure*}

Figure~\ref{fig:colcol} shows the \emph{g-r} and \emph{r-i} diagram for BCGs (red), SHs (black), and ICL (blue) for each individual system in our sample, as a function of redshift (different panels). We can see that there is a diversity of colours, spanning a wide range in both cases. Overall, it appears that all three components have similar colours, and independently of redshift, but also all of them progressively redden from high redshift to the present day. At $z=0$, we compare our predictions with the observed SH colours from VEGAS (green diamonds for VG and brown squares for FDS). The uncertainties in the observational data are broad enough to encompass most of our predicted values, although the distribution appears somewhat more scattered. Nevertheless, the majority of the observed points fall within the cloud of model predictions. This is visible by eye when comparing the black triangle with error bars (indicating mean and $\pm 1\sigma$ distribution of the predicted SH colours). Once we consider $\pm 2\sigma$, most of the green diamonds and error bars fall within the predicted distribution.

In order to catch any possible difference between the colours of the different components, we plot in Figure~\ref{fig:coldistr} their distributions at $z=0$ (left panel) and $z=2$ (right panel). Solid lines represent
the \emph{g-r}, while the dashed ones indicate \emph{r-i}, and line colours denote the three components as in the previous figure. Plotted together with our predictions at $z=0$, we show the mean values of the two observed colours (thick green solid and dashed lines) along with their corresponding $\pm 1\sigma$ distributions (thin green solid and dashed lines).

There are several key points worth discussing, mainly in the difference between the two redshifts investigated. First of all, as already noted in Figure~\ref{fig:colcol}, both colours tend to become redder with time, which is a natural consequence of ageing. For SHs and ICL it is understandable since they do not have any star formation, and ageing overcomes the newly stripped stars that might be blue. For the BCGs instead, it just means that they age by passive evolution with little episodes of star formation, after a period of growth due to mergers (e.g., \citealt{oliva-altamirano2014,lee2017} or \citealt{contini2024e} for a review), and the fact that \emph{g-r} grows faster than \emph{r-i} is a clear indication of that.

Another important feature of the plot is that at $z=0$ all components and for both colours, have narrower distributions than those at $z=2$, and with more pronounced peaks. This highlights the variety of the populations at high redshifts with respect to an old population at the present day. Not only, at $z=0$ the distributions appear to be alike, while that is not the case at $z=2$. At this redshift, when the Universe was just 3 Gyr old, BCGs are consistently redder than the other two components. This trend is seen even at $z=0$, but in a lesser degree, meaning that SHs and ICL age faster than BCGs. Again, this is understandable given the fact that star formation is not happening in these two components. More importantly, the model predictions for both colours show excellent agreement with the observed SH colours.

To summarize, our model predicts a negative gradient in colours from the inner component, the BCG, to the outer component, the ICL, but this trend weakens over time. Moreover, there is not statistical distinction between SHs and ICL, implying that they have very similar colours at any time. Once again, this can be explained by the tight link between the components, given that SHs form directly from stars originally belonging to the ICL.

Taken together, our results indicate that broadband colours alone are insufficient to disentangle SHs from the surrounding ICL: the two components display virtually indistinguishable \emph{g-r} and \emph{r-i} distributions at all epochs, while BCGs remain only mildly redder on average by $z=0$. This degeneracy is physically expected if SHs are continuously replenished by the same accreted populations that build the ICL, and it is consistent with the large object-to-object scatter in observed colour gradients reported for nearby halos and extended envelopes \citep[e.g.][]{merritt2016,gilhuly2022}. In the Local Group, both the Milky Way and M31 show negative colour/metallicity gradients in their outskirts, but with amplitudes that strongly depend on each system’s recent accretion history \citep{kalirai2006,gilbert2014,deason2014}. Cosmological simulations likewise predict broadly similar colours for ex–situ dominated components—outer BCG envelopes, SHs, and ICL—modulated by the stochastic contribution of a few massive progenitors \citep{pillepich2018,monachesi2019,elias2020,wright2024}. Practically, our findings imply that future wide-field programs (e.g. LSST imaging combined with WEAVE/4MOST spectroscopy) should rely on chemo-dynamical tracers rather than broadband colours alone to unambiguously separate SHs from the diffuse intracluster population \citep{cooper2011,helmi2020,spavone2020,spavone2021,spavone2022,iodice2016}.

\begin{figure*}[t!]
\begin{center}
\begin{tabular}{cc}
\includegraphics[width=0.47\textwidth]{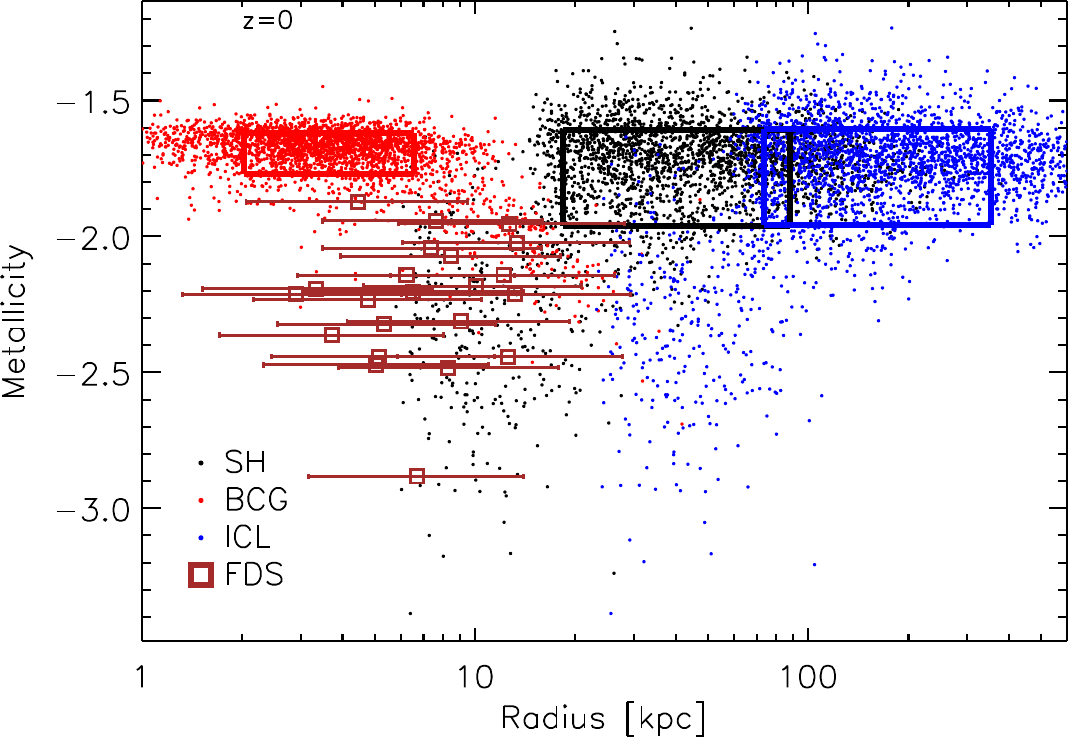} &
\includegraphics[width=0.47\textwidth]{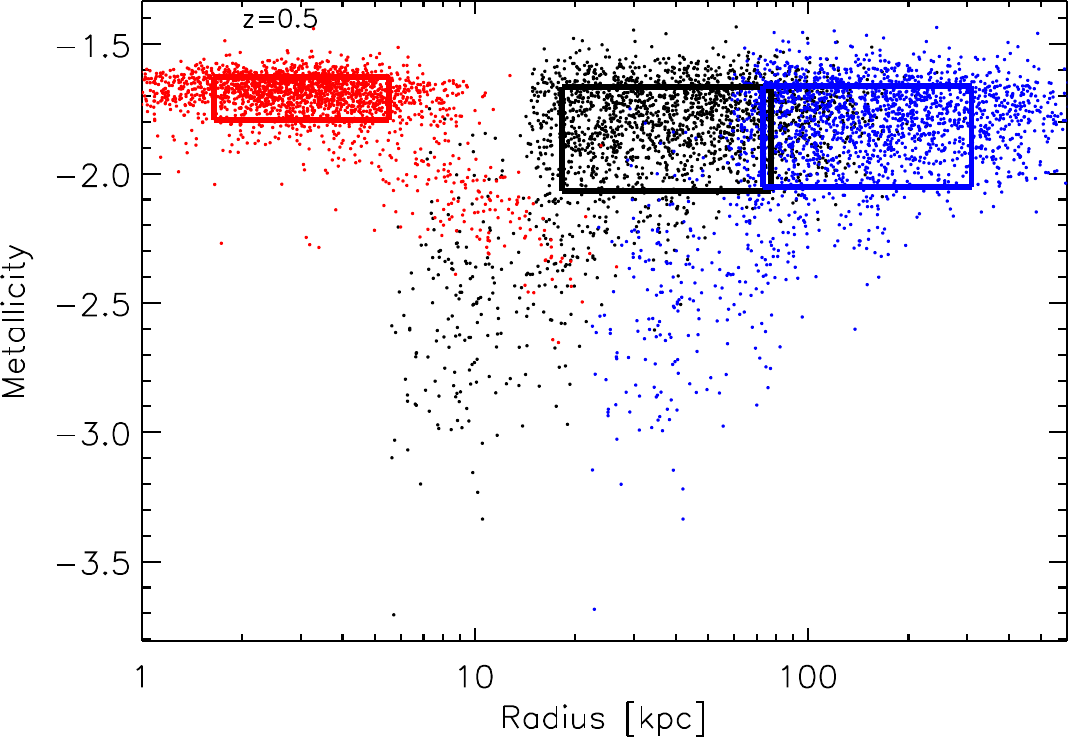} \\
\includegraphics[width=0.47\textwidth]{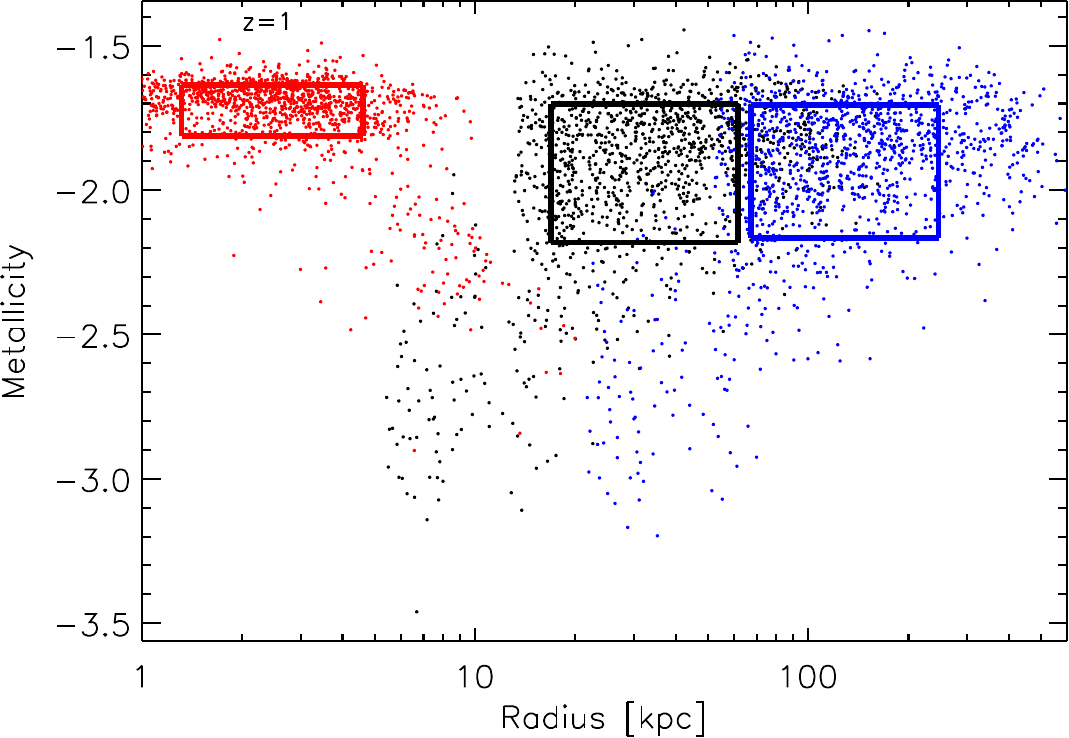} &
\includegraphics[width=0.47\textwidth]{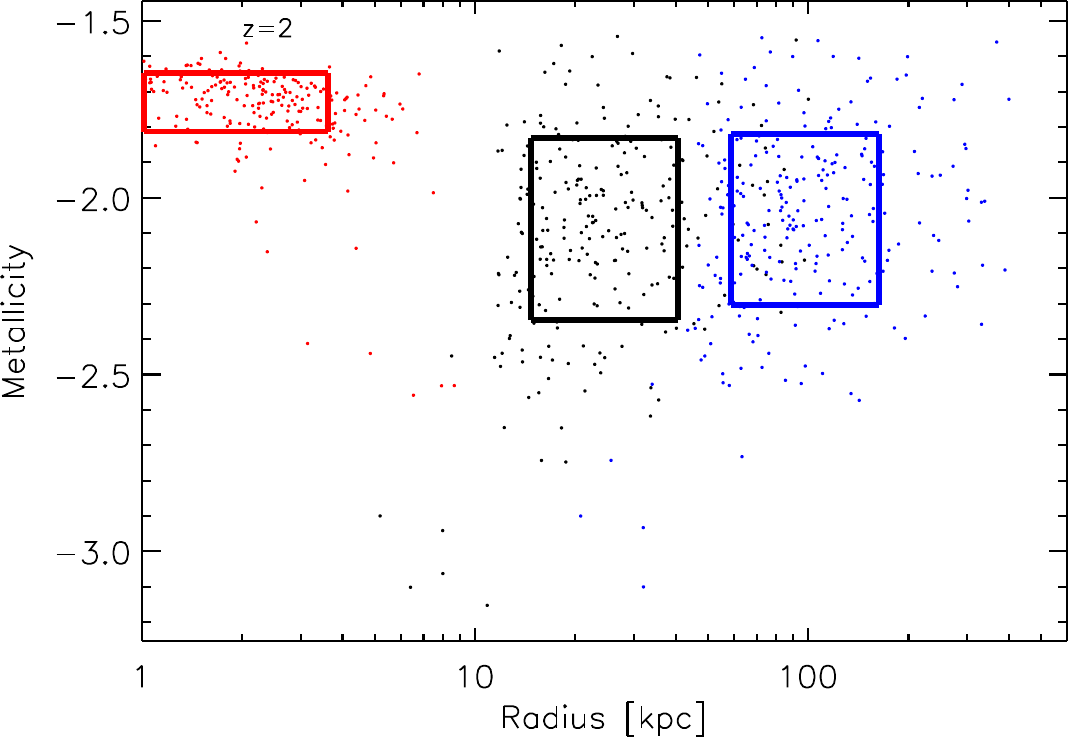} \\
\end{tabular}
\caption{Metallicity profiles ($\log Z$) of the three components as a function of radius (see text for details), shown at different redshifts (separate panels). At high redshift, both stellar halos and the ICL tend to be more metal-poor than the BCGs, although this trend becomes less evident at lower redshifts. The shaded rectangles mark the regions enclosing the 16th to 84th percentiles of the distributions, while brown squares represent the observed metallicities in FDS at $z=0$.}
\label{fig:metradius}
\end{center}
\end{figure*}

By focusing on the metallicity of the three components, we can assess whether the trends observed for colours are also present in the metallicity distributions. In Figure~\ref{fig:metradius}, we plot the metallicity of the three components of each system in our sample, as a function of their radius and at the usual redshifts (different panels). The coloured rectangles show the 16th and 84th distribution for each component. The predictions at $z=0$ are accompanied by observed SH metallicities from FDS (brown squares) only, because for VG they are not available. Before delving into the features of the plot, it is worth explaining how we calculate the radius of each component, in order to place them in the plot. In the case of the BCG, its radius is defined as done in \citet{contini2019}, mass-weighting the bulge and disk radii, i.e.:
\begin{equation}\label{eqn:rbcg}
 R_{\rm BCG} = \frac{R_{\rm bulge} \cdot M_{\rm bulge}^* +1.68\cdot R_{\rm sl}\cdot M_{\rm disk}^*}{M_{\rm BCG}^*} \, ,
\end{equation}
where $R_{\rm bulge}$ and $1.68\cdot R_{\rm sl}$ are the half stellar mass radii of the bulge and disk, $M_{\rm bulge}^*$, $M_{\rm disk}^*$ and $M_{\rm BCG}^*$ are the stellar masses of the bulge, disk and BCG as a whole, respectively.
SHs are defined in terms of the transition radius $R_{\rm trans}$; accordingly, we place them at half of this radius. Similarly, for FDS data we place them at $R_{2}/2$. For the ICL instead, there is no clear placement because, in principle, they can extend up to the virial radius. Considering that the "real" radius has no physical impact in the context of this plot, we place the ICL to $2\cdot R_{\rm trans}$.

The plot shows us that both SHs and ICL are more metal poor than the BCGs at high redshift, but the gap becomes smaller approaching redshift $z=0$, where the difference in metallicity between the three components is
indistinguishable. Not only, SHs and ICL have the same metallicity independently of the redshift. FDS data lie within the cloud of predicted SH metallicities on the left side of the panel, closely tracing the BCG distribution. This behavior likely arises from two factors: first, the observationally defined $R_2$ differs intrinsically from our model definition of $R_{\rm trans}$, though the two are physically related; second, while our analysis focuses on central galaxies in the SAM, the FDS sample mainly consists of satellites, which are on average less massive than the model centrals.

\begin{figure*}[t!]
\begin{center}
\begin{tabular}{cc}
\includegraphics[width=0.47\textwidth]{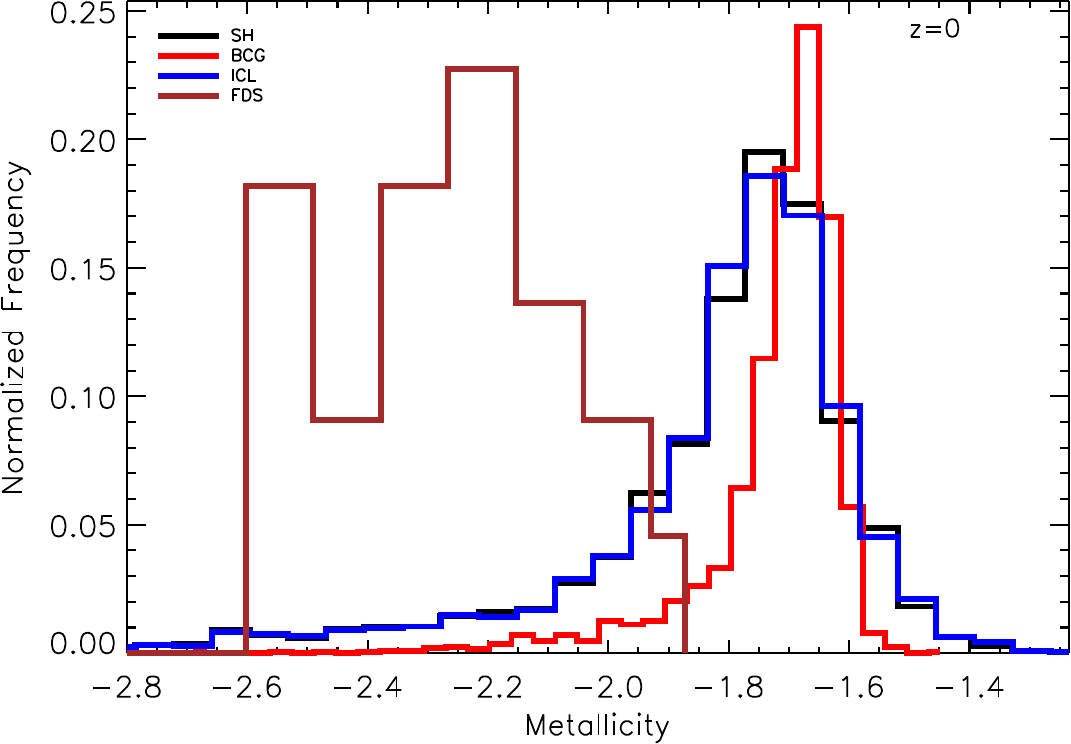} &
\includegraphics[width=0.47\textwidth]{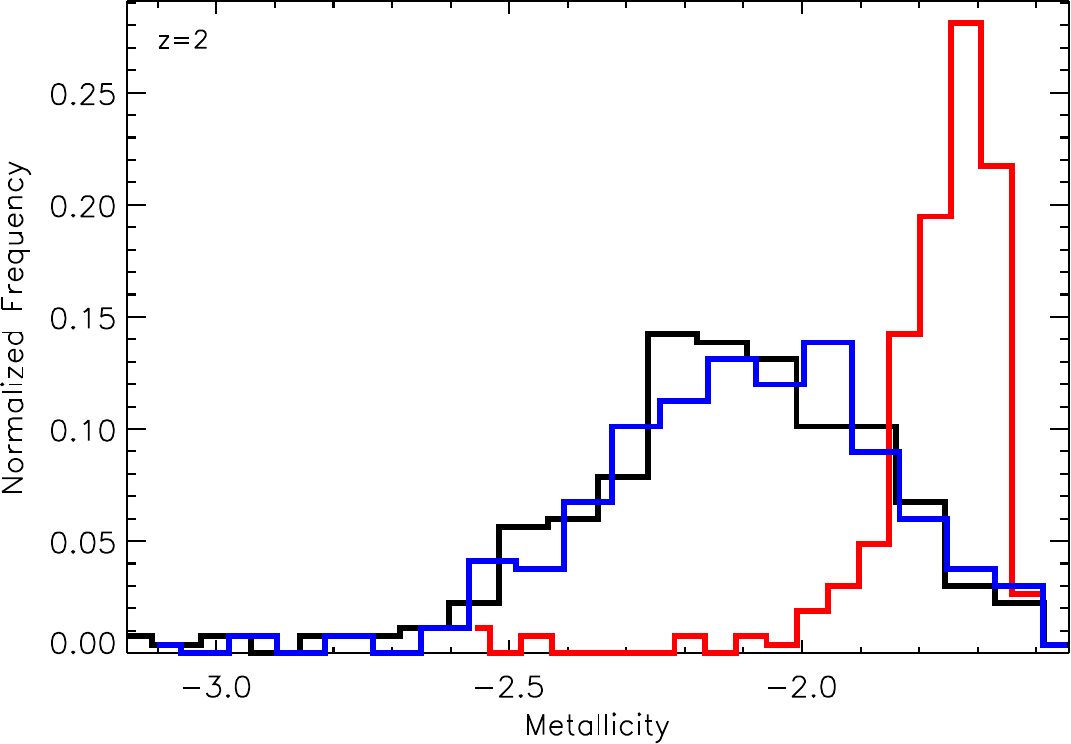} \\
\end{tabular}
\caption{Similarly to Figure~\ref{fig:coldistr}, the plots display the metallicity distributions ($\log Z$) for BCGs (red), stellar halos (black), and ICL (blue) at $z=0$ (left panel) and $z=2$ (right panel). In both panels, stellar halos and the ICL exhibit broadly comparable distributions, whereas BCGs remain systematically more metal-rich than the other two components, regardless of redshift. The separation is present at all epochs, but appears much more pronounced at $z=2$. The brown line at $z=0$ represent the observed distribution in FDS (SHs).}
\label{fig:metdistr}
\end{center}
\end{figure*}

As done above with colours, we need to quantify in a better way the real difference, if any. So, we plot in Figure~\ref{fig:metdistr} the metallicity distributions of BCGs, SHs and ICL, at $z=0$ (left panel) and $z=2$ (right panel). By focusing on $z=2$, we can see that, while the metallicity distributions of SHs and ICL are very similar, almost identical, that of BCGs is very much different and peaks on higher metallicity. The peak for the BCGs is at $\sim -1.7$, while those of SHs and ICL is wide, between $\sim -2.0$ and $\sim -2.3$, on average at $-2.15$. Hence, the gap in metallicity between BCGs and SH-ICL is around 0.4 dex at $z=2$.

At $z=0$, the picture is rather different. The three distributions are very similar, although those of SHs and ICL are wider. Considering the peaks, the average metallicity of BCGs does not change with respect to that at $z=2$ ($\sim -1.7$), but SHs and ICL become more metal rich as time goes by, almost reaching the average metallicity of BCGs. The gap between the mean metallicity of BCGs and that of SHs-ICL is around 0.1 dex. As noted above—and therefore expected—the FDS SH metallicity distribution peaks at systematically lower values than our predictions, suggesting that these SHs likely originate from disrupted dwarf galaxies, as noticed in Section~\ref{sec:intro}.

To summarize, similarly to colours, our model predicts a negative gradient from the BCG to the ICL, which is very clear at high redshifts, but almost flattens over time. Again, as for the colours, SH and ICL have the same metallicity.

The redshift evolution of metallicity strengthens the above picture. At $z=2$ we find a $\simeq0.4$ dex offset between BCGs and the SH–ICL pair, with the latter two being essentially indistinguishable; by $z=0$ the gap shrinks to $\simeq0.1$ dex as SHs and the ICL become more metal rich. This convergence naturally arises if late-time growth in the outskirts is increasingly driven by the stripping of relatively massive, chemically evolved satellites, while early assembly is dominated by lower-mass, metal-poor dwarfs. The trend echoes simulation results in which ex–situ accretion builds extended BCG envelopes at late times and predicts shallow, slowly evolving outer metallicity gradients \citep{nelson2024,montenegro-taborda2023,contreras-santos2024,pillepich2018}. Conversely, the systematically lower metallicities measured in our observed samples are consistent with a larger fractional contribution from disrupted dwarfs \citep[e.g.][]{mouhcine2005,harmsen2017,dsouza2018,spavone2017,ragusa2021,ragusa2022}, plausibly reflecting environmental or selection differences with respect to our central-galaxy sample. In this context, the average metallicity of SHs/ICL is a sensitive diagnostic of the mass spectrum of destroyed progenitors; combining metallicity with $\alpha$–enhancement and kinematics should further discriminate rapid early enrichment from extended, low-mass accretion histories \citep{venn2004,deason2016}.

\FloatBarrier
\section{Conclusions}\label{sec:conclusions}

In this work we have investigated the properties of stellar halos (SHs) of bright
central galaxies (BCGs), building on our previous implementation of SH formation
in the semianalytic model \textsc{FEGA25} and extending the analysis to their
redshift evolution in terms of scaling relations, colours, and metallicities. Our
main findings can be summarized as follows:

\paragraph{1. Scaling with halo structure.}
The SH mass correlates tightly with both the BCG and intracluster light (ICL)
masses. While both relations are basically linear, the SH--ICL relation exhibits
significantly smaller scatter, as expected from the adopted SH
definition, consistent with the direct origin of SHs from ICL stars. Halo
concentration emerges as a fundamental driver: less concentrated (typically more
massive) halos host more extended transition radii and correspondingly larger
SHs. In addition, SHs are more massive in systems with higher ICL formation
efficiency, showing that both structural and baryonic processes shape the
SH--ICL connection. We note that the residual scatter in these relations
reflects variations in halo concentration, ICL assembly histories, and
physically motivated stability criteria applied in the model.

\paragraph{2. Transition radii.}
The distribution of $R_{\rm trans}$ predicted by the model peaks around
30--40~kpc nearly independently of redshift, with $\sim$90\% of systems below
100~kpc. However, massive halos can reach $R_{\rm trans}\sim 400$~kpc at $z=0$.
The weak redshift evolution suggests that SH sizes are largely set by halo
concentration rather than epoch, in agreement with recent
simulation-based determinations (e.g.\citealt{proctor2024}).
At the same time, the observational $R_{\rm trans}$ distribution inferred
from VEGAS is more narrowly peaked at smaller radii, indicating an offset with
respect to the model predictions. This difference reflects the distinct
definitions adopted in observations and in the semi-analytic model, as well as
sample-selection effects. The comparison should therefore be interpreted in a
statistical and qualitative sense, rather than as a direct one-to-one
correspondence between individual systems.

\paragraph{3. Colors.}
All three components---BCGs, SHs, and ICL---progressively redden with cosmic
time. By $z=0$, their colour distributions largely overlap, with only a mild
offset leaving BCGs redder on average. SHs and ICL are virtually indistinguishable
in broad-band colours at all epochs, implying that colours alone are not sufficient
to separate these two components observationally. This degeneracy reinforces the
idea of a common physical origin and is consistent with the large scatter of
observed colour gradients in nearby halos. In the model, colours are
derived self-consistently from the star-formation histories of the different
stellar components, rather than treated as free parameters.

\paragraph{4. Metallicities.}
At $z=2$, BCGs are more metal rich than both SHs and ICL by $\sim$0.4~dex, but
this gap shrinks to $\sim$0.1~dex by $z=0$ as SHs/ICL become progressively more
enriched through the stripping of more massive satellites. The convergence of
metallicities over time highlights the role of late accretion in building
chemically evolved outskirts. FDS observations, however, peak at lower
metallicities, suggesting that disrupted dwarf galaxies contribute more strongly
to observed halos than in our central-galaxy sample. This points to possible
environmental or selection effects and emphasizes the diagnostic power of
metallicity gradients for constraining progenitor mass functions.

\paragraph{5. Overall picture.}
SHs appear as transition regions between the stars bound to BCGs and those
belonging to the diffuse ICL. We emphasize that this interpretation
follows directly from the adopted SH definition, while the trends in mass,
structure, and stellar populations discussed in this work reflect genuine
physical dependencies on halo concentration, ICL formation efficiency, and the
mass spectrum of accreted satellites. These conclusions support a view in which
SHs are not isolated components but rather the inner manifestation of the ICL,
dynamically and chemically coupled to the outer galaxy environment. Future
wide-field imaging and spectroscopic surveys (e.g.\ LSST, WEAVE, 4MOST) will be
crucial to test these predictions by simultaneously probing structure,
metallicity, and kinematics across large samples of halos in diverse
environments.

\begin{acknowledgements}
The authors thank the anonymous referee for their constructive comments, which have significantly contributed to improving this manuscript.
E.C. and S.K.Y. acknowledge support from the Korean National Research Foundation (RS-2025-00514475; RS-2022-NR070872), and E.C. acknowledges support from the Korean National Research Foundation (RS-2023-00241934). M.S. and E.I. acknowledge the support by the Italian Ministry for 1224 Education University and Research (MIUR) grant PRIN 2022 2022383WFT 1225
“SUNRISE”, CUP C53D23000850006 and by VST funds. R.R. acknowledges financial support through grants PRIN-MIUR 2020SKSTHZ and through INAF-WEAVE StePS funds.
\end{acknowledgements}

 \bibliographystyle{aa.bst}
  \bibliography{biblio}

@ARTICLE{devaucouleurs1959,
       author = {{de Vaucouleurs}, G.},
        title = "{Classification and Morphology of External Galaxies}",
      journal = {Handbuch der Physik},
         year = 1959,
       volume = {53},
        pages = {275},
       adsurl = {https://ui.adsabs.harvard.edu/abs/1959HDP....53..275D}
}

@ARTICLE{gallart2019,
       author = {Gallart, C. and Bernard, E.~J. and Brook, C.~B. and others},
        title = "{Uncovering the birth of the Milky Way through chemodynamics}",
      journal = {Nature Astronomy},
         year = 2019,
        month = sep,
       volume = {3},
        pages = {932-939},
          doi = {10.1038/s41550-019-0829-5},
archivePrefix = {arXiv},
       eprint = {1901.02900},
 primaryClass = {astro-ph.GA},
       adsurl = {https://ui.adsabs.harvard.edu/abs/2019NatAs...3..932G}
}

@ARTICLE{sarzi2018,
       author = {{Sarzi}, M. and {Iodice}, E. and {Coccato}, L. and {Corsini}, E.~M. and {de Zeeuw}, P.~T. and {Falc{\'o}n-Barroso}, J. and {Gadotti}, D.~A. and {Lyubenova}, M. and {McDermid}, R.~M. and {van de Ven}, G. and {Fahrion}, K. and {Pizzella}, A. and {Zhu}, L.},
        title = "{Fornax3D project: Overall goals, galaxy sample, MUSE data analysis, and initial results}",
      journal = {\aap},
         year = 2018,
        month = aug,
       volume = {616},
          eid = {A121},
        pages = {A121},
          doi = {10.1051/0004-6361/201833137},
archivePrefix = {arXiv},
       eprint = {1804.06795},
 primaryClass = {astro-ph.GA},
       adsurl = {https://ui.adsabs.harvard.edu/abs/2018A&A...616A.121S}
}

@ARTICLE{krajnovic2018,
       author = {{Krajnovi{\'c}}, Davor and {Emsellem}, Eric and {den Brok}, Mark and {Marino}, Raffaella Anna and {Schmidt}, Kasper Borello and {Steinmetz}, Matthias and {Weilbacher}, Peter M.},
        title = "{Climbing to the top of the galactic mass ladder: evidence for frequent prolate-like rotation among the most massive galaxies}",
      journal = {\mnras},
         year = 2018,
        month = jul,
       volume = {477},
       number = {4},
        pages = {5327-5337},
          doi = {10.1093/mnras/sty1031},
archivePrefix = {arXiv},
       eprint = {1802.02591},
 primaryClass = {astro-ph.GA},
       adsurl = {https://ui.adsabs.harvard.edu/abs/2018MNRAS.477.5327K}
}

@ARTICLE{naidu2020,
       author = {Naidu, R.~P. and Conroy, C. and Bonaca, A. and others},
        title = "{The Splash: Substructure in the Inner Halo of the Milky Way}",
      journal = {\apj},
         year = 2020,
        month = sep,
       volume = {901},
       number = {1},
        pages = {48},
          doi = {10.3847/1538-4357/abaef4},
archivePrefix = {arXiv},
       eprint = {2006.08625},
 primaryClass = {astro-ph.GA},
       adsurl = {https://ui.adsabs.harvard.edu/abs/2020ApJ...901...48N}
}

@ARTICLE{amorisco2017,
       author = {{Amorisco}, N.~C.},
        title = "{Contributions to the accreted stellar halo: an atlas of stellar deposition}",
      journal = {\mnras},
         year = 2017,
        month = jan,
       volume = {464},
       number = {3},
        pages = {2882-2895},
          doi = {10.1093/mnras/stw2229},
archivePrefix = {arXiv},
       eprint = {1511.08806},
 primaryClass = {astro-ph.GA},
       adsurl = {https://ui.adsabs.harvard.edu/abs/2017MNRAS.464.2882A}
}

@ARTICLE{bahe2017,
       author = {{Bah{\'e}}, Yannick M. and {Barnes}, David J. and {Dalla Vecchia}, Claudio and {Kay}, Scott T. and {White}, Simon D.~M. and {McCarthy}, Ian G. and {Schaye}, Joop and {Bower}, Richard G. and {Crain}, Robert A. and {Theuns}, Tom and {Jenkins}, Adrian and {McGee}, Sean L. and {Schaller}, Matthieu and {Thomas}, Peter A. and {Trayford}, James W.},
        title = "{The Hydrangea simulations: galaxy formation in and around massive clusters}",
      journal = {\mnras},
         year = 2017,
        month = oct,
       volume = {470},
       number = {4},
        pages = {4186-4208},
          doi = {10.1093/mnras/stx1403},
archivePrefix = {arXiv},
       eprint = {1703.10610},
 primaryClass = {astro-ph.GA},
       adsurl = {https://ui.adsabs.harvard.edu/abs/2017MNRAS.470.4186B}
}

@ARTICLE{barnes2017,
       author = {{Barnes}, David J. and {Kay}, Scott T. and {Bah{\'e}}, Yannick M. and {Dalla Vecchia}, Claudio and {McCarthy}, Ian G. and {Schaye}, Joop and {Bower}, Richard G. and {Jenkins}, Adrian and {Thomas}, Peter A. and {Schaller}, Matthieu and {Crain}, Robert A. and {Theuns}, Tom and {White}, Simon D.~M.},
        title = "{The Cluster-EAGLE project: global properties of simulated clusters with resolved galaxies}",
      journal = {\mnras},
         year = 2017,
        month = oct,
       volume = {471},
       number = {1},
        pages = {1088-1106},
          doi = {10.1093/mnras/stx1647},
archivePrefix = {arXiv},
       eprint = {1703.10907},
 primaryClass = {astro-ph.GA},
       adsurl = {https://ui.adsabs.harvard.edu/abs/2017MNRAS.471.1088B}
}

@ARTICLE{beers2012,
       author = {{Beers}, Timothy C. and {Carollo}, Daniela and {Ivezi{\'c}}, {\v{Z}}eljko and {An}, Deokkeun and {Chiba}, Masashi and {Norris}, John E. and {Freeman}, Ken C. and {Lee}, Young Sun and {Munn}, Jeffrey A. and {Re Fiorentin}, Paola and {Sivarani}, Thirupathi and {Wilhelm}, Ronald and {Yanny}, Brian and {York}, Donald G.},
        title = "{The Case for the Dual Halo of the Milky Way}",
      journal = {\apj},
         year = 2012,
        month = feb,
       volume = {746},
       number = {1},
          eid = {34},
        pages = {34},
          doi = {10.1088/0004-637X/746/1/34},
archivePrefix = {arXiv},
       eprint = {1104.2513},
 primaryClass = {astro-ph.GA},
       adsurl = {https://ui.adsabs.harvard.edu/abs/2012ApJ...746...34B}
}

@ARTICLE{belokurov2018,
       author = {{Belokurov}, V. and {Erkal}, D. and {Evans}, N.~W. and {Koposov}, S.~E. and {Deason}, A.~J.},
        title = "{Co-formation of the disc and the stellar halo}",
      journal = {\mnras},
         year = 2018,
        month = jul,
       volume = {478},
       number = {1},
        pages = {611-619},
          doi = {10.1093/mnras/sty982},
archivePrefix = {arXiv},
       eprint = {1802.03414},
 primaryClass = {astro-ph.GA},
       adsurl = {https://ui.adsabs.harvard.edu/abs/2018MNRAS.478..611B}
}

@ARTICLE{belokurov2020,
       author = {Belokurov, V. and Sanders, J.~L. and Fattahi, A. and others},
        title = "{Co-formation of the disc and the stellar halo}",
      journal = {\mnras},
         year = 2020,
        month = may,
       volume = {494},
        pages = {3880-3898},
          doi = {10.1093/mnras/staa1015},
archivePrefix = {arXiv},
       eprint = {1909.04679},
 primaryClass = {astro-ph.GA},
       adsurl = {https://ui.adsabs.harvard.edu/abs/2020MNRAS.494.3880B}
}

@ARTICLE{beltrand2024,
       author = {{Beltrand}, Camila and {Monachesi}, Antonela and {D'Souza}, Richard and {Bell}, Eric F. and {de Jong}, Roelof S. and {Gomez}, Facundo A. and {Bailin}, Jeremy and {Jang}, In Sung and {Smercina}, Adam},
        title = "{First resolved stellar halo kinematics of a Milky Way-mass galaxy outside the Local Group: The flat counter-rotating halo in NGC 4945}",
      journal = {\aap},
         year = 2024,
        month = oct,
       volume = {690},
          eid = {A115},
        pages = {A115},
          doi = {10.1051/0004-6361/202450626},
archivePrefix = {arXiv},
       eprint = {2406.17533},
 primaryClass = {astro-ph.GA},
       adsurl = {https://ui.adsabs.harvard.edu/abs/2024A&A...690A.115B}
}

@ARTICLE{bullock2005,
       author = {{Bullock}, James S. and {Johnston}, Kathryn V.},
        title = "{Tracing Galaxy Formation with Stellar Halos. I. Methods}",
      journal = {\apj},
         year = 2005,
        month = dec,
       volume = {635},
       number = {2},
        pages = {931-949},
          doi = {10.1086/497422},
archivePrefix = {arXiv},
       eprint = {astro-ph/0506467},
 primaryClass = {astro-ph},
       adsurl = {https://ui.adsabs.harvard.edu/abs/2005ApJ...635..931B}
}

@ARTICLE{capaccioli2015,
       author = {{Capaccioli}, Massimo and {Spavone}, Marilena and {Grado}, Aniello and {Iodice}, Enrichetta and {Limatola}, Luca and {Napolitano}, Nicola R. and {Cantiello}, Michele and {Paolillo}, Maurizio and {Romanowsky}, Aaron J. and {Forbes}, Duncan A. and {Puzia}, Thomas H. and {Raimondo}, Gabriella and {Schipani}, Pietro},
        title = "{VEGAS: A VST Early-type GAlaxy Survey. I. Presentation, wide-field surface photometry, and substructures in NGC 4472}",
      journal = {\aap},
         year = 2015,
        month = sep,
       volume = {581},
          eid = {A10},
        pages = {A10},
          doi = {10.1051/0004-6361/201526252},
archivePrefix = {arXiv},
       eprint = {1507.01336},
 primaryClass = {astro-ph.GA},
       adsurl = {https://ui.adsabs.harvard.edu/abs/2015A&A...581A..10C}
}

@ARTICLE{carollo2007,
       author = {{Carollo}, Daniela and {Beers}, Timothy C. and {Lee}, Young Sun and {Chiba}, Masashi and {Norris}, John E. and {Wilhelm}, Ronald and {Sivarani}, Thirupathi and {Marsteller}, Brian and {Munn}, Jeffrey A. and {Bailer-Jones}, Coryn A.~L. and {Fiorentin}, Paola Re and {York}, Donald G.},
        title = "{Two stellar components in the halo of the Milky Way}",
      journal = {\nat},
         year = 2007,
        month = dec,
       volume = {450},
       number = {7172},
        pages = {1020-1025},
          doi = {10.1038/nature06460},
archivePrefix = {arXiv},
       eprint = {0706.3005},
 primaryClass = {astro-ph},
       adsurl = {https://ui.adsabs.harvard.edu/abs/2007Natur.450.1020C}
}

@ARTICLE{carollo2010,
       author = {Carollo, D. and Beers, T.~C. and Chiba, M. and others},
        title = "{Structure and Kinematics of the Milky Way Halo}",
      journal = {\apj},
         year = 2010,
        month = mar,
       volume = {712},
       number = {1},
        pages = {692-727},
          doi = {10.1088/0004-637X/712/1/692},
archivePrefix = {arXiv},
       eprint = {0909.3019},
 primaryClass = {astro-ph.GA},
       adsurl = {https://ui.adsabs.harvard.edu/abs/2010ApJ...712..692C}
}

@ARTICLE{chabrier2003,
       author = {{Chabrier}, Gilles},
        title = "{Galactic Stellar and Substellar Initial Mass Function}",
      journal = {\pasp},
         year = 2003,
        month = jul,
       volume = {115},
       number = {809},
        pages = {763-795},
          doi = {10.1086/376392},
archivePrefix = {arXiv},
       eprint = {astro-ph/0304382},
 primaryClass = {astro-ph},
       adsurl = {https://ui.adsabs.harvard.edu/abs/2003PASP..115..763C}
}

@ARTICLE{child2018,
       author = {{Child}, Hillary L. and {Habib}, Salman and {Heitmann}, Katrin and {Frontiere}, Nicholas and {Finkel}, Hal and {Pope}, Adrian and {Morozov}, Vitali},
        title = "{Halo Profiles and the Concentration-Mass Relation for a {\ensuremath{\Lambda}}CDM Universe}",
      journal = {\apj},
         year = 2018,
        month = may,
       volume = {859},
       number = {1},
          eid = {55},
        pages = {55},
          doi = {10.3847/1538-4357/aabf95},
archivePrefix = {arXiv},
       eprint = {1804.10199},
 primaryClass = {astro-ph.CO},
       adsurl = {https://ui.adsabs.harvard.edu/abs/2018ApJ...859...55C}
}

@ARTICLE{contini2014,
       author = {{Contini}, E. and {De Lucia}, G. and {Villalobos}, {\'A}. and {Borgani}, S.},
        title = "{On the formation and physical properties of the intracluster light in hierarchical galaxy formation models}",
      journal = {\mnras},
         year = 2014,
        month = feb,
       volume = {437},
       number = {4},
        pages = {3787-3802},
          doi = {10.1093/mnras/stt2174},
archivePrefix = {arXiv},
       eprint = {1311.2076},
 primaryClass = {astro-ph.CO},
       adsurl = {https://ui.adsabs.harvard.edu/abs/2014MNRAS.437.3787C}
}

@ARTICLE{contini2018,
       author = {{Contini}, E. and {Yi}, S.~K. and {Kang}, X.},
        title = "{The different growth pathways of brightest cluster galaxies and intracluster light}",
      journal = {\mnras},
         year = 2018,
        month = sep,
       volume = {479},
       number = {1},
        pages = {932-944},
          doi = {10.1093/mnras/sty1518},
archivePrefix = {arXiv},
       eprint = {1806.01480},
 primaryClass = {astro-ph.GA},
       adsurl = {https://ui.adsabs.harvard.edu/abs/2018MNRAS.479..932C}
}

@ARTICLE{contini2019,
       author = {{Contini}, E. and {Yi}, S.~K. and {Kang}, X.},
        title = "{Theoretical Predictions of Colors and Metallicity of the Intracluster Light}",
      journal = {\apj},
         year = 2019,
        month = jan,
       volume = {871},
       number = {1},
          eid = {24},
        pages = {24},
          doi = {10.3847/1538-4357/aaf41f},
archivePrefix = {arXiv},
       eprint = {1811.03253},
 primaryClass = {astro-ph.GA},
       adsurl = {https://ui.adsabs.harvard.edu/abs/2019ApJ...871...24C}
}

@ARTICLE{contini2020b,
       author = {{Contini}, E. and {Gu}, Q.},
        title = "{On the Mass Distribution of the Intracluster Light in Galaxy Groups and Clusters}",
      journal = {\apj},
         year = 2020,
        month = oct,
       volume = {901},
       number = {2},
          eid = {128},
        pages = {128},
          doi = {10.3847/1538-4357/abb1aa},
archivePrefix = {arXiv},
       eprint = {2005.13763},
 primaryClass = {astro-ph.GA},
       adsurl = {https://ui.adsabs.harvard.edu/abs/2020ApJ...901..128C}
}

@ARTICLE{contini2023,
       author = {{Contini}, Emanuele and {Jeon}, Seyoung and {Rhee}, Jinsu and {Han}, San and {Yi}, Sukyoung K.},
        title = "{The Intracluster Light and Its Link with the Dynamical State of the Host Group/Cluster: The Role of the Halo Concentration}",
      journal = {\apj},
         year = 2023,
        month = nov,
       volume = {958},
       number = {1},
          eid = {72},
        pages = {72},
          doi = {10.3847/1538-4357/acfd25},
archivePrefix = {arXiv},
       eprint = {2310.03263},
 primaryClass = {astro-ph.GA},
       adsurl = {https://ui.adsabs.harvard.edu/abs/2023ApJ...958...72C}
}

@ARTICLE{contini2024,
       author = {{Contini}, Emanuele and {Rhee}, Jinsu and {Han}, San and {Jeon}, Seyoung and {Yi}, Sukyoung K.},
        title = "{The Connection between the Intracluster Light and its Host Halo: Formation Time and Contribution from Different Channels}",
      journal = {\aj},
         year = 2024,
        month = jan,
       volume = {167},
       number = {1},
          eid = {7},
        pages = {7},
          doi = {10.3847/1538-3881/ad0894},
archivePrefix = {arXiv},
       eprint = {2310.20135},
 primaryClass = {astro-ph.GA},
       adsurl = {https://ui.adsabs.harvard.edu/abs/2024AJ....167....7C}
}

@ARTICLE{contini2024c,
       author = {{Contini}, Emanuele and {Yi}, Sukyoung K. and {Jeon}, Seyoung and {Rhee}, Jinsu},
        title = "{The Impact of Positive AGN Feedback on the Properties of Galaxies in a Semianalytic Model of Galaxy Formation}",
      journal = {\apjs},
         year = 2024,
        month = oct,
       volume = {274},
       number = {2},
          eid = {41},
        pages = {41},
          doi = {10.3847/1538-4365/ad70ac},
       adsurl = {https://ui.adsabs.harvard.edu/abs/2024ApJS..274...41C}
}

@ARTICLE{contini2024d,
       author = {{Contini}, Emanuele and {Spavone}, Marilena and {Ragusa}, Rossella and {Iodice}, Enrichetta and {Yi}, Sukyoung K.},
        title = "{Stellar halos of bright central galaxies: A view from the FEGA semi-analytic model of galaxy formation and VEGAS survey}",
      journal = {\aap},
         year = 2024,
        month = dec,
       volume = {692},
          eid = {L9},
        pages = {L9},
          doi = {10.1051/0004-6361/202452741},
archivePrefix = {arXiv},
       eprint = {2411.10663},
 primaryClass = {astro-ph.GA},
       adsurl = {https://ui.adsabs.harvard.edu/abs/2024A&A...692L...9C}
}

@ARTICLE{contini2024e,
       author = {{Contini}, Emanuele and {Yi}, Sukyoung K. and {Jeon}, Seyoung},
        title = "{Brightest Cluster Galaxies and the Intracluster Light}",
      journal = {arXiv e-prints},
         year = 2024,
        month = apr,
          eid = {arXiv:2404.01560},
        pages = {arXiv:2404.01560},
          doi = {10.48550/arXiv.2404.01560},
archivePrefix = {arXiv},
       eprint = {2404.01560},
 primaryClass = {astro-ph.GA},
       adsurl = {https://ui.adsabs.harvard.edu/abs/2024arXiv240401560C}
}

@ARTICLE{contini2025a,
       author = {{Contini}, Emanuele and {Yi}, Sukyoung K. and {Rhee}, Jinsu and {Jeon}, Seyoung},
        title = "{A Full Active Galactic Nuclei Feedback Prescription for Numerical Models: Negative, Positive, and Hot Gas-ejection Modes}",
      journal = {\apjs},
         year = 2025,
        month = jul,
       volume = {279},
       number = {1},
          eid = {18},
        pages = {18},
          doi = {10.3847/1538-4365/addb4b},
archivePrefix = {arXiv},
       eprint = {2502.19503},
 primaryClass = {astro-ph.GA},
       adsurl = {https://ui.adsabs.harvard.edu/abs/2025ApJS..279...18C}
}

@ARTICLE{contini2025b,
       author = {{Contini}, Emanuele and {Seo}, Changjo and {Rhee}, Jinsu and {Jeon}, Seyoung and {Yi}, Sukyoung K.},
        title = "{Roles of Supernova and Active Galactic Nucleus Feedback in Shaping the Baryonic Content in a Wide Range of Dark Matter Halo Masses}",
      journal = {\apjs},
         year = 2025,
        month = nov,
       volume = {281},
       number = {1},
          eid = {2},
        pages = {2},
          doi = {10.3847/1538-4365/ae060f},
archivePrefix = {arXiv},
       eprint = {2507.10673},
 primaryClass = {astro-ph.GA},
       adsurl = {https://ui.adsabs.harvard.edu/abs/2025ApJS..281....2C}
}

@ARTICLE{contreras-santos2024,
       author = {{Contreras-Santos}, A. and {Knebe}, A. and {Cui}, W. and {Alonso Asensio}, I. and {Dalla Vecchia}, C. and {Ca{\~n}as}, R. and {Haggar}, R. and {Mostoghiu Paun}, R.~A. and {Pearce}, F.~R. and {Rasia}, E.},
        title = "{Characterising the intra-cluster light in The Three Hundred simulations}",
      journal = {\aap},
         year = 2024,
        month = mar,
       volume = {683},
          eid = {A59},
        pages = {A59},
          doi = {10.1051/0004-6361/202348474},
archivePrefix = {arXiv},
       eprint = {2401.08283},
 primaryClass = {astro-ph.CO},
       adsurl = {https://ui.adsabs.harvard.edu/abs/2024A&A...683A..59C}
}

@ARTICLE{cooper2010,
       author = {{Cooper}, A.~P. and {Cole}, S. and {Frenk}, C.~S. and {White}, S.~D.~M. and {Helly}, J. and {Benson}, A.~J. and {De Lucia}, G. and {Helmi}, A. and {Jenkins}, A. and {Navarro}, J.~F. and {Springel}, V. and {Wang}, J.},
        title = "{Galactic stellar haloes in the CDM model}",
      journal = {\mnras},
         year = 2010,
        month = aug,
       volume = {406},
       number = {2},
        pages = {744-766},
          doi = {10.1111/j.1365-2966.2010.16740.x},
archivePrefix = {arXiv},
       eprint = {0910.3211},
 primaryClass = {astro-ph.GA},
       adsurl = {https://ui.adsabs.harvard.edu/abs/2010MNRAS.406..744C}
}

@ARTICLE{cooper2011,
       author = {{Cooper}, Andrew P. and {Mart{\'\i}nez-Delgado}, David and {Helly}, John and {Frenk}, Carlos and {Cole}, Shaun and {Crawford}, Ken and {Zibetti}, Stefano and {Carballo-Bello}, Julio A. and {GaBany}, R. Jay},
        title = "{The Formation of Shell Galaxies Similar to NGC 7600 in the Cold Dark Matter Cosmogony}",
      journal = {\apjl},
         year = 2011,
        month = dec,
       volume = {743},
       number = {1},
          eid = {L21},
        pages = {L21},
          doi = {10.1088/2041-8205/743/1/L21},
archivePrefix = {arXiv},
       eprint = {1111.2864},
 primaryClass = {astro-ph.GA},
       adsurl = {https://ui.adsabs.harvard.edu/abs/2011ApJ...743L..21C}
}

@ARTICLE{cooper2015,
       author = {{Cooper}, Andrew P. and {Parry}, Owen H. and {Lowing}, Ben and {Cole}, Shaun and {Frenk}, Carlos},
        title = "{Formation of in situ stellar haloes in Milky Way-mass galaxies}",
      journal = {\mnras},
         year = 2015,
        month = dec,
       volume = {454},
       number = {3},
        pages = {3185-3199},
          doi = {10.1093/mnras/stv2057},
archivePrefix = {arXiv},
       eprint = {1501.04630},
 primaryClass = {astro-ph.GA},
       adsurl = {https://ui.adsabs.harvard.edu/abs/2015MNRAS.454.3185C}
}

@ARTICLE{correa2015,
       author = {{Correa}, Camila A. and {Wyithe}, J. Stuart B. and {Schaye}, Joop and {Duffy}, Alan R.},
        title = "{The accretion history of dark matter haloes - III. A physical model for the concentration-mass relation}",
      journal = {\mnras},
         year = 2015,
        month = sep,
       volume = {452},
       number = {2},
        pages = {1217-1232},
          doi = {10.1093/mnras/stv1363},
archivePrefix = {arXiv},
       eprint = {1502.00391},
 primaryClass = {astro-ph.CO},
       adsurl = {https://ui.adsabs.harvard.edu/abs/2015MNRAS.452.1217C}
}

@ARTICLE{crain2015,
       author = {{Crain}, Robert A. and {Schaye}, Joop and {Bower}, Richard G. and {Furlong}, Michelle and {Schaller}, Matthieu and {Theuns}, Tom and {Dalla Vecchia}, Claudio and {Frenk}, Carlos S. and {McCarthy}, Ian G. and {Helly}, John C. and {Jenkins}, Adrian and {Rosas-Guevara}, Yetli M. and {White}, Simon D.~M. and {Trayford}, James W.},
        title = "{The EAGLE simulations of galaxy formation: calibration of subgrid physics and model variations}",
      journal = {\mnras},
         year = 2015,
        month = jun,
       volume = {450},
       number = {2},
        pages = {1937-1961},
          doi = {10.1093/mnras/stv725},
archivePrefix = {arXiv},
       eprint = {1501.01311},
 primaryClass = {astro-ph.GA},
       adsurl = {https://ui.adsabs.harvard.edu/abs/2015MNRAS.450.1937C}
}

@ARTICLE{cui2018,
       author = {{Cui}, Weiguang and {Knebe}, Alexander and {Yepes}, Gustavo and {Pearce}, Frazer and {Power}, Chris and {Dave}, Romeel and {Arth}, Alexander and {Borgani}, Stefano and {Dolag}, Klaus and {Elahi}, Pascal and {Mostoghiu}, Robert and {Murante}, Giuseppe and {Rasia}, Elena and {Stoppacher}, Doris and {Vega-Ferrero}, Jesus and {Wang}, Yang and {Yang}, Xiaohu and {Benson}, Andrew and {Cora}, Sof{\'\i}a A. and {Croton}, Darren J. and {Sinha}, Manodeep and {Stevens}, Adam R.~H. and {Vega-Mart{\'\i}nez}, Cristian A. and {Arthur}, Jake and {Baldi}, Anna S. and {Ca{\~n}as}, Rodrigo and {Cialone}, Giammarco and {Cunnama}, Daniel and {De Petris}, Marco and {Durando}, Giacomo and {Ettori}, Stefano and {Gottl{\"o}ber}, Stefan and {Nuza}, Sebasti{\'a}n E. and {Old}, Lyndsay J. and {Pilipenko}, Sergey and {Sorce}, Jenny G. and {Welker}, Charlotte},
        title = "{The Three Hundred project: a large catalogue of theoretically modelled galaxy clusters for cosmological and astrophysical applications}",
      journal = {\mnras},
         year = 2018,
        month = nov,
       volume = {480},
       number = {3},
        pages = {2898-2915},
          doi = {10.1093/mnras/sty2111},
archivePrefix = {arXiv},
       eprint = {1809.04622},
 primaryClass = {astro-ph.GA},
       adsurl = {https://ui.adsabs.harvard.edu/abs/2018MNRAS.480.2898C}
}

@ARTICLE{deason2013,
       author = {Deason, A.~J. and Belokurov, V. and Evans, N.~W.},
        title = "{Accreted and In Situ Halo Stars in the Milky Way}",
      journal = {\apj},
         year = 2013,
        month = feb,
       volume = {763},
       number = {2},
        pages = {113},
          doi = {10.1088/0004-637X/763/2/113},
archivePrefix = {arXiv},
       eprint = {1210.5505},
 primaryClass = {astro-ph.GA},
       adsurl = {https://ui.adsabs.harvard.edu/abs/2013ApJ...763..113D}
}

@ARTICLE{deason2014,
       author = {{Deason}, A.~J. and {Belokurov}, V. and {Koposov}, S.~E. and {Rockosi}, C.~M.},
        title = "{Touching The Void: A Striking Drop in Stellar Halo Density Beyond 50 kpc}",
      journal = {\apj},
         year = 2014,
        month = may,
       volume = {787},
       number = {1},
          eid = {30},
        pages = {30},
          doi = {10.1088/0004-637X/787/1/30},
archivePrefix = {arXiv},
       eprint = {1403.7205},
 primaryClass = {astro-ph.GA},
       adsurl = {https://ui.adsabs.harvard.edu/abs/2014ApJ...787...30D}
}

@ARTICLE{deason2016,
       author = {{Deason}, Alis J. and {Mao}, Yao-Yuan and {Wechsler}, Risa H.},
        title = "{The Eating Habits of Milky Way-mass Halos: Destroyed Dwarf Satellites and the Metallicity Distribution of Accreted Stars}",
      journal = {\apj},
         year = 2016,
        month = apr,
       volume = {821},
       number = {1},
          eid = {5},
        pages = {5},
          doi = {10.3847/0004-637X/821/1/5},
archivePrefix = {arXiv},
       eprint = {1601.07905},
 primaryClass = {astro-ph.GA},
       adsurl = {https://ui.adsabs.harvard.edu/abs/2016ApJ...821....5D}
}

@ARTICLE{dejong2019,
       author = {{de Jong}, R.~S. and {Agertz}, O. and {Berbel}, A.~A. and {Aird}, J. and {Alexander}, D.~A. and {Amarsi}, A. and {Anders}, F. and {Andrae}, R. and {Ansarinejad}, B. and {Ansorge}, W. and {Antilogus}, P. and {Anwand-Heerwart}, H. and {Arentsen}, A. and {Arnadottir}, A. and {Asplund}, M. and {Auger}, M. and {Azais}, N. and {Baade}, D. and {Baker}, G. and {Baker}, S. and {Balbinot}, E. and {Baldry}, I.~K. and {Banerji}, M. and {Barden}, S. and {Barklem}, P. and {Barth{\'e}l{\'e}my-Mazot}, E. and {Battistini}, C. and {Bauer}, S. and {Bell}, C.~P.~M. and {Bellido-Tirado}, O. and {Bellstedt}, S. and {Belokurov}, V. and {Bensby}, T. and {Bergemann}, M. and {Bestenlehner}, J.~M. and {Bielby}, R. and {Bilicki}, M. and {Blake}, C. and {Bland-Hawthorn}, J. and {Boeche}, C. and {Boland}, W. and {Boller}, T. and {Bongard}, S. and {Bongiorno}, A. and {Bonifacio}, P. and {Boudon}, D. and {Brooks}, D. and {Brown}, M.~J.~I. and {Brown}, R. and {Br{\"u}ggen}, M. and {Brynnel}, J. and {Brzeski}, J. and {Buchert}, T. and {Buschkamp}, P. and {Caffau}, E. and {Caillier}, P. and {Carrick}, J. and {Casagrande}, L. and {Case}, S. and {Casey}, A. and {Cesarini}, I. and {Cescutti}, G. and {Chapuis}, D. and {Chiappini}, C. and {Childress}, M. and {Christlieb}, N. and {Church}, R. and {Cioni}, M. -R.~L. and {Cluver}, M. and {Colless}, M. and {Collett}, T. and {Comparat}, J. and {Cooper}, A. and {Couch}, W. and {Courbin}, F. and {Croom}, S. and {Croton}, D. and {Daguis{\'e}}, E. and {Dalton}, G. and {Davies}, L.~J.~M. and {Davis}, T. and {de Laverny}, P. and {Deason}, A. and {Dionies}, F. and {Disseau}, K. and {Doel}, P. and {D{\"o}scher}, D. and {Driver}, S.~P. and {Dwelly}, T. and {Eckert}, D. and {Edge}, A. and {Edvardsson}, B. and {Youssoufi}, D.~E. and {Elhaddad}, A. and {Enke}, H. and {Erfanianfar}, G. and {Farrell}, T. and {Fechner}, T. and {Feiz}, C. and {Feltzing}, S. and {Ferreras}, I. and {Feuerstein}, D. and {Feuillet}, D. and {Finoguenov}, A. and {Ford}, D. and {Fotopoulou}, S. and {Fouesneau}, M. and {Frenk}, C. and {Frey}, S. and {Gaessler}, W. and {Geier}, S. and {Gentile Fusillo}, N. and {Gerhard}, O. and {Giannantonio}, T. and {Giannone}, D. and {Gibson}, B. and {Gillingham}, P. and {Gonz{\'a}lez-Fern{\'a}ndez}, C. and {Gonzalez-Solares}, E. and {Gottloeber}, S. and {Gould}, A. and {Grebel}, E.~K. and {Gueguen}, A. and {Guiglion}, G. and {Haehnelt}, M. and {Hahn}, T. and {Hansen}, C.~J. and {Hartman}, H. and {Hauptner}, K. and {Hawkins}, K. and {Haynes}, D. and {Haynes}, R. and {Heiter}, U. and {Helmi}, A. and {Aguayo}, C.~H. and {Hewett}, P. and {Hinton}, S. and {Hobbs}, D. and {Hoenig}, S. and {Hofman}, D. and {Hook}, I. and {Hopgood}, J. and {Hopkins}, A. and {Hourihane}, A. and {Howes}, L. and {Howlett}, C. and {Huet}, T. and {Irwin}, M. and {Iwert}, O. and {Jablonka}, P. and {Jahn}, T. and {Jahnke}, K. and {Jarno}, A. and {Jin}, S. and {Jofre}, P. and {Johl}, D. and {Jones}, D. and {J{\"o}nsson}, H. and {Jordan}, C. and {Karovicova}, I. and {Khalatyan}, A. and {Kelz}, A. and {Kennicutt}, R. and {King}, D. and {Kitaura}, F. and {Klar}, J. and {Klauser}, U. and {Kneib}, J. -P. and {Koch}, A. and {Koposov}, S. and {Kordopatis}, G. and {Korn}, A. and {Kosmalski}, J. and {Kotak}, R. and {Kovalev}, M. and {Kreckel}, K. and {Kripak}, Y. and {Krumpe}, M. and {Kuijken}, K. and {Kunder}, A. and {Kushniruk}, I. and {Lam}, M.~I. and {Lamer}, G. and {Laurent}, F. and {Lawrence}, J. and {Lehmitz}, M. and {Lemasle}, B. and {Lewis}, J. and {Li}, B. and {Lidman}, C. and {Lind}, K. and {Liske}, J. and {Lizon}, J. -L. and {Loveday}, J. and {Ludwig}, H. -G. and {McDermid}, R.~M. and {Maguire}, K. and {Mainieri}, V. and {Mali}, S. and {Mandel}, H.},
        title = "{4MOST: Project overview and information for the First Call for Proposals}",
      journal = {The Messenger},
         year = 2019,
        month = mar,
       volume = {175},
        pages = {3-11},
          doi = {10.18727/0722-6691/5117},
archivePrefix = {arXiv},
       eprint = {1903.02464},
 primaryClass = {astro-ph.IM},
       adsurl = {https://ui.adsabs.harvard.edu/abs/2019Msngr.175....3D}
}

@ARTICLE{drinkwater2001,
       author = {{Drinkwater}, Michael J. and {Gregg}, Michael D. and {Colless}, Matthew},
        title = "{Substructure and Dynamics of the Fornax Cluster}",
      journal = {\apjl},
         year = 2001,
        month = feb,
       volume = {548},
       number = {2},
        pages = {L139-L142},
          doi = {10.1086/319113},
archivePrefix = {arXiv},
       eprint = {astro-ph/0012415},
 primaryClass = {astro-ph},
       adsurl = {https://ui.adsabs.harvard.edu/abs/2001ApJ...548L.139D}
}

@ARTICLE{dsouza2018,
       author = {{D'Souza}, Richard and {Bell}, Eric F.},
        title = "{The masses and metallicities of stellar haloes reflect galactic merger histories}",
      journal = {\mnras},
         year = 2018,
        month = mar,
       volume = {474},
       number = {4},
        pages = {5300-5318},
          doi = {10.1093/mnras/stx3081},
archivePrefix = {arXiv},
       eprint = {1705.08442},
 primaryClass = {astro-ph.GA},
       adsurl = {https://ui.adsabs.harvard.edu/abs/2018MNRAS.474.5300D}
}

@ARTICLE{duc2015,
       author = {{Duc}, Pierre-Alain and {Cuillandre}, Jean-Charles and {Karabal}, Emin and {Cappellari}, Michele and {Alatalo}, Katherine and {Blitz}, Leo and {Bournaud}, Fr{\'e}d{\'e}ric and {Bureau}, Martin and {Crocker}, Alison F. and {Davies}, Roger L. and {Davis}, Timothy A. and {de Zeeuw}, P.~T. and {Emsellem}, Eric and {Khochfar}, Sadegh and {Krajnovi{\'c}}, Davor and {Kuntschner}, Harald and {McDermid}, Richard M. and {Michel-Dansac}, Leo and {Morganti}, Raffaella and {Naab}, Thorsten and {Oosterloo}, Tom and {Paudel}, Sanjaya and {Sarzi}, Marc and {Scott}, Nicholas and {Serra}, Paolo and {Weijmans}, Anne-Marie and {Young}, Lisa M.},
        title = "{The ATLAS$^{3D}$ project - XXIX. The new look of early-type galaxies and surrounding fields disclosed by extremely deep optical images}",
      journal = {\mnras},
         year = 2015,
        month = jan,
       volume = {446},
       number = {1},
        pages = {120-143},
          doi = {10.1093/mnras/stu2019},
archivePrefix = {arXiv},
       eprint = {1410.0981},
 primaryClass = {astro-ph.GA},
       adsurl = {https://ui.adsabs.harvard.edu/abs/2015MNRAS.446..120D}
}

@article{efron1979,
  title        = {Bootstrap methods: Another look at the jackknife},
  author       = {Efron, Bradley},
  journal      = {The Annals of Statistics},
  volume       = {7},
  number       = {1},
  pages        = {1--26},
  year         = {1979},
  publisher    = {Institute of Mathematical Statistics},
  doi          = {10.1214/aos/1176344552}
}

@ARTICLE{elias2020,
       author = {{Elias}, Lydia M. and {Sales}, Laura V. and {Helmi}, Amina and {Hernquist}, Lars},
        title = "{Cosmological insights into the assembly of the radial and compact stellar halo of the Milky Way}",
      journal = {\mnras},
         year = 2020,
        month = jun,
       volume = {495},
       number = {1},
        pages = {29-39},
          doi = {10.1093/mnras/staa1090},
archivePrefix = {arXiv},
       eprint = {2003.03381},
 primaryClass = {astro-ph.GA},
       adsurl = {https://ui.adsabs.harvard.edu/abs/2020MNRAS.495...29E}
}

@ARTICLE{font2011,
       author = {{Font}, A.~S. and {McCarthy}, I.~G. and {Crain}, R.~A. and {Theuns}, T. and {Schaye}, J. and {Wiersma}, R.~P.~C. and {Dalla Vecchia}, C.},
        title = "{Cosmological simulations of the formation of the stellar haloes around disc galaxies}",
      journal = {\mnras},
         year = 2011,
        month = oct,
       volume = {416},
       number = {4},
        pages = {2802-2820},
          doi = {10.1111/j.1365-2966.2011.19227.x},
archivePrefix = {arXiv},
       eprint = {1102.2526},
 primaryClass = {astro-ph.CO},
       adsurl = {https://ui.adsabs.harvard.edu/abs/2011MNRAS.416.2802F}
}

@ARTICLE{gendron2025,
       author = {{Gendron}, Val{\'e}rie and {Martel}, Hugo},
        title = "{Numerical simulations of starbursts triggered by tidal disruption in clusters: searching for observational signatures}",
      journal = {\mnras},
         year = 2025,
        month = aug,
       volume = {541},
       number = {3},
        pages = {2513-2539},
          doi = {10.1093/mnras/staf1142},
       adsurl = {https://ui.adsabs.harvard.edu/abs/2025MNRAS.541.2513G}
}

@ARTICLE{gilbert2014,
       author = {{Gilbert}, Karoline M. and {Kalirai}, Jason S. and {Guhathakurta}, Puragra and {Beaton}, Rachael L. and {Geha}, Marla C. and {Kirby}, Evan N. and {Majewski}, Steven R. and {Patterson}, Richard J. and {Tollerud}, Erik J. and {Bullock}, James S. and {Tanaka}, Mikito and {Chiba}, Masashi},
        title = "{Global Properties of M31's Stellar Halo from the SPLASH Survey. II. Metallicity Profile}",
      journal = {\apj},
         year = 2014,
        month = dec,
       volume = {796},
       number = {2},
          eid = {76},
        pages = {76},
          doi = {10.1088/0004-637X/796/2/76},
archivePrefix = {arXiv},
       eprint = {1409.3843},
 primaryClass = {astro-ph.GA},
       adsurl = {https://ui.adsabs.harvard.edu/abs/2014ApJ...796...76G}
}

@ARTICLE{gilhuly2022,
       author = {{Gilhuly}, Colleen and {Merritt}, Allison and {Abraham}, Roberto and {Danieli}, Shany and {Lokhorst}, Deborah and {Liu}, Qing and {van Dokkum}, Pieter and {Conroy}, Charlie and {Greco}, Johnny},
        title = "{Stellar Halos from the The Dragonfly Edge-on Galaxies Survey}",
      journal = {\apj},
         year = 2022,
        month = jun,
       volume = {932},
       number = {1},
          eid = {44},
        pages = {44},
          doi = {10.3847/1538-4357/ac6750},
archivePrefix = {arXiv},
       eprint = {2204.06596},
 primaryClass = {astro-ph.GA},
       adsurl = {https://ui.adsabs.harvard.edu/abs/2022ApJ...932...44G}
}

@ARTICLE{grand2017,
       author = {{Grand}, Robert J.~J. and {G{\'o}mez}, Facundo A. and {Marinacci}, Federico and {Pakmor}, R{\"u}diger and {Springel}, Volker and {Campbell}, David J.~R. and {Frenk}, Carlos S. and {Jenkins}, Adrian and {White}, Simon D.~M.},
        title = "{The Auriga Project: the properties and formation mechanisms of disc galaxies across cosmic time}",
      journal = {\mnras},
         year = 2017,
        month = may,
       volume = {467},
       number = {1},
        pages = {179-207},
          doi = {10.1093/mnras/stx071},
archivePrefix = {arXiv},
       eprint = {1610.01159},
 primaryClass = {astro-ph.GA},
       adsurl = {https://ui.adsabs.harvard.edu/abs/2017MNRAS.467..179G}
}

@ARTICLE{guo2011,
       author = {{Guo}, Qi and {White}, Simon and {Boylan-Kolchin}, Michael and {De Lucia}, Gabriella and {Kauffmann}, Guinevere and {Lemson}, Gerard and {Li}, Cheng and {Springel}, Volker and {Weinmann}, Simone},
        title = "{From dwarf spheroidals to cD galaxies: simulating the galaxy population in a {\ensuremath{\Lambda}}CDM cosmology}",
      journal = {\mnras},
         year = 2011,
        month = may,
       volume = {413},
       number = {1},
        pages = {101-131},
          doi = {10.1111/j.1365-2966.2010.18114.x},
archivePrefix = {arXiv},
       eprint = {1006.0106},
 primaryClass = {astro-ph.CO},
       adsurl = {https://ui.adsabs.harvard.edu/abs/2011MNRAS.413..101G}
}

@ARTICLE{harmsen2017,
       author = {{Harmsen}, Benjamin and {Monachesi}, Antonela and {Bell}, Eric F. and {de Jong}, Roelof S. and {Bailin}, Jeremy and {Radburn-Smith}, David J. and {Holwerda}, Benne W.},
        title = "{Diverse stellar haloes in nearby Milky Way mass disc galaxies}",
      journal = {\mnras},
         year = 2017,
        month = apr,
       volume = {466},
       number = {2},
        pages = {1491-1512},
          doi = {10.1093/mnras/stw2992},
archivePrefix = {arXiv},
       eprint = {1611.05448},
 primaryClass = {astro-ph.GA},
       adsurl = {https://ui.adsabs.harvard.edu/abs/2017MNRAS.466.1491H}
}

@ARTICLE{helmi2008,
       author = {{Helmi}, Amina},
        title = "{The stellar halo of the Galaxy}",
      journal = {\aapr},
         year = 2008,
        month = jun,
       volume = {15},
       number = {3},
        pages = {145-188},
          doi = {10.1007/s00159-008-0009-6},
archivePrefix = {arXiv},
       eprint = {0804.0019},
 primaryClass = {astro-ph},
       adsurl = {https://ui.adsabs.harvard.edu/abs/2008A&ARv..15..145H}
}

@ARTICLE{helmi2018,
       author = {{Helmi}, Amina and {Babusiaux}, Carine and {Koppelman}, Helmer H. and {Massari}, Davide and {Veljanoski}, Jovan and {Brown}, Anthony G.~A.},
        title = "{The merger that led to the formation of the Milky Way's inner stellar halo and thick disk}",
      journal = {\nat},
         year = 2018,
        month = oct,
       volume = {563},
       number = {7729},
        pages = {85-88},
          doi = {10.1038/s41586-018-0625-x},
archivePrefix = {arXiv},
       eprint = {1806.06038},
 primaryClass = {astro-ph.GA},
       adsurl = {https://ui.adsabs.harvard.edu/abs/2018Natur.563...85H}
}

@ARTICLE{helmi2020,
       author = {{Helmi}, Amina},
        title = "{Streams, Substructures, and the Early History of the Milky Way}",
      journal = {\araa},
         year = 2020,
        month = aug,
       volume = {58},
        pages = {205-256},
          doi = {10.1146/annurev-astro-032620-021917},
archivePrefix = {arXiv},
       eprint = {2002.04340},
 primaryClass = {astro-ph.GA},
       adsurl = {https://ui.adsabs.harvard.edu/abs/2020ARA&A..58..205H}
}

@ARTICLE{hopkins2014,
       author = {{Hopkins}, Philip F. and {Kere{\v{s}}}, Du{\v{s}}an and {O{\~n}orbe}, Jos{\'e} and {Faucher-Gigu{\`e}re}, Claude-Andr{\'e} and {Quataert}, Eliot and {Murray}, Norman and {Bullock}, James S.},
        title = "{Galaxies on FIRE (Feedback In Realistic Environments): stellar feedback explains cosmologically inefficient star formation}",
      journal = {\mnras},
         year = 2014,
        month = nov,
       volume = {445},
       number = {1},
        pages = {581-603},
          doi = {10.1093/mnras/stu1738},
archivePrefix = {arXiv},
       eprint = {1311.2073},
 primaryClass = {astro-ph.CO},
       adsurl = {https://ui.adsabs.harvard.edu/abs/2014MNRAS.445..581H}
}

@ARTICLE{ibata2014,
       author = {{Ibata}, Rodrigo A. and {Lewis}, Geraint F. and {McConnachie}, Alan W. and {Martin}, Nicolas F. and {Irwin}, Michael J. and {Ferguson}, Annette M.~N. and {Babul}, Arif and {Bernard}, Edouard J. and {Chapman}, Scott C. and {Collins}, Michelle and {Fardal}, Mark and {Mackey}, A.~D. and {Navarro}, Julio and {Pe{\~n}arrubia}, Jorge and {Rich}, R. Michael and {Tanvir}, Nial and {Widrow}, Lawrence},
        title = "{The Large-scale Structure of the Halo of the Andromeda Galaxy. I. Global Stellar Density, Morphology and Metallicity Properties}",
      journal = {\apj},
         year = 2014,
        month = jan,
       volume = {780},
       number = {2},
          eid = {128},
        pages = {128},
          doi = {10.1088/0004-637X/780/2/128},
archivePrefix = {arXiv},
       eprint = {1311.5888},
 primaryClass = {astro-ph.GA},
       adsurl = {https://ui.adsabs.harvard.edu/abs/2014ApJ...780..128I}
}

@ARTICLE{iodice2016,
       author = {{Iodice}, E. and {Capaccioli}, M. and {Grado}, A. and {Limatola}, L. and {Spavone}, M. and {Napolitano}, N.~R. and {Paolillo}, M. and {Peletier}, R.~F. and {Cantiello}, M. and {Lisker}, T. and {Wittmann}, C. and {Venhola}, A. and {Hilker}, M. and {D'Abrusco}, R. and {Pota}, V. and {Schipani}, P.},
        title = "{The Fornax Deep Survey with VST. I. The Extended and Diffuse Stellar Halo of NGC 1399 out to 192 kpc}",
      journal = {\apj},
         year = 2016,
        month = mar,
       volume = {820},
       number = {1},
          eid = {42},
        pages = {42},
          doi = {10.3847/0004-637X/820/1/42},
archivePrefix = {arXiv},
       eprint = {1602.02149},
 primaryClass = {astro-ph.GA},
       adsurl = {https://ui.adsabs.harvard.edu/abs/2016ApJ...820...42I}
}

@ARTICLE{iodice2017,
       author = {{Iodice}, E. and {Spavone}, M. and {Cantiello}, M. and {D'Abrusco}, R. and {Capaccioli}, M. and {Hilker}, M. and {Mieske}, S. and {Napolitano}, N.~R. and {Peletier}, R.~F. and {Limatola}, L. and {Grado}, A. and {Venhola}, A. and {Paolillo}, M. and {Van de Ven}, G. and {Schipani}, P.},
        title = "{Intracluster Patches of Baryons in the Core of the Fornax Cluster}",
      journal = {\apj},
         year = 2017,
        month = dec,
       volume = {851},
       number = {2},
          eid = {75},
        pages = {75},
          doi = {10.3847/1538-4357/aa9b30},
archivePrefix = {arXiv},
       eprint = {1711.04681},
 primaryClass = {astro-ph.GA},
       adsurl = {https://ui.adsabs.harvard.edu/abs/2017ApJ...851...75I}
}

@ARTICLE{iodice2019,
       author = {{Iodice}, E. and {Spavone}, M. and {Capaccioli}, M. and {Peletier}, R.~F. and {van de Ven}, G. and {Napolitano}, N.~R. and {Hilker}, M. and {Mieske}, S. and {Smith}, R. and {Pasquali}, A. and {Limatola}, L. and {Grado}, A. and {Venhola}, A. and {Cantiello}, M. and {Paolillo}, M. and {Falcon-Barroso}, J. and {D'Abrusco}, R. and {Schipani}, P.},
        title = "{The Fornax Deep Survey with the VST. V. Exploring the faintest regions of the bright early-type galaxies inside the virial radius}",
      journal = {\aap},
         year = 2019,
        month = mar,
       volume = {623},
          eid = {A1},
        pages = {A1},
          doi = {10.1051/0004-6361/201833741},
archivePrefix = {arXiv},
       eprint = {1812.01050},
 primaryClass = {astro-ph.GA},
       adsurl = {https://ui.adsabs.harvard.edu/abs/2019A&A...623A...1I}
}

@ARTICLE{iodice2021,
       author = {{Iodice}, E. and {Spavone}, M. and {Capaccioli}, M. and {Schipani}, P. and {Arnaboldi}, M. and {Cantiello}, M. and {D'Ago}, G. and {De Cicco}, D. and {Forbes}, D.~A. and {Greggio}, L. and {Krajnovi{\'c}}, D. and {La Marca}, A. and {Napolitano}, N.~R. and {Paolillo}, M. and {Ragusa}, R. and {Raj}, M.~A. and {Rampazzo}, R. and {Rejkuba}, M.},
        title = "{The VST Early-type GAlaxy Survey: Exploring the Outskirts and Intra-cluster Regions of Galaxies in the Low-surface- brightness Regime}",
      journal = {The Messenger},
         year = 2021,
        month = jun,
       volume = {183},
        pages = {25-29},
          doi = {10.18727/0722-6691/5232},
archivePrefix = {arXiv},
       eprint = {2106.06448},
 primaryClass = {astro-ph.GA},
       adsurl = {https://ui.adsabs.harvard.edu/abs/2021Msngr.183...25I}
}

@ARTICLE{ivezic2019,
       author = {{Ivezi{\'c}}, {\v{Z}}eljko and {Kahn}, Steven M. and {Tyson}, J. Anthony and {Abel}, Bob and {Acosta}, Emily and {Allsman}, Robyn and {Alonso}, David and {AlSayyad}, Yusra and {Anderson}, Scott F. and {Andrew}, John and {Angel}, James Roger P. and {Angeli}, George Z. and {Ansari}, Reza and {Antilogus}, Pierre and {Araujo}, Constanza and {Armstrong}, Robert and {Arndt}, Kirk T. and {Astier}, Pierre and {Aubourg}, {\'E}ric and {Auza}, Nicole and {Axelrod}, Tim S. and {Bard}, Deborah J. and {Barr}, Jeff D. and {Barrau}, Aurelian and {Bartlett}, James G. and {Bauer}, Amanda E. and {Bauman}, Brian J. and {Baumont}, Sylvain and {Bechtol}, Ellen and {Bechtol}, Keith and {Becker}, Andrew C. and {Becla}, Jacek and {Beldica}, Cristina and {Bellavia}, Steve and {Bianco}, Federica B. and {Biswas}, Rahul and {Blanc}, Guillaume and {Blazek}, Jonathan and {Blandford}, Roger D. and {Bloom}, Josh S. and {Bogart}, Joanne and {Bond}, Tim W. and {Booth}, Michael T. and {Borgland}, Anders W. and {Borne}, Kirk and {Bosch}, James F. and {Boutigny}, Dominique and {Brackett}, Craig A. and {Bradshaw}, Andrew and {Brandt}, William Nielsen and {Brown}, Michael E. and {Bullock}, James S. and {Burchat}, Patricia and {Burke}, David L. and {Cagnoli}, Gianpietro and {Calabrese}, Daniel and {Callahan}, Shawn and {Callen}, Alice L. and {Carlin}, Jeffrey L. and {Carlson}, Erin L. and {Chandrasekharan}, Srinivasan and {Charles-Emerson}, Glenaver and {Chesley}, Steve and {Cheu}, Elliott C. and {Chiang}, Hsin-Fang and {Chiang}, James and {Chirino}, Carol and {Chow}, Derek and {Ciardi}, David R. and {Claver}, Charles F. and {Cohen-Tanugi}, Johann and {Cockrum}, Joseph J. and {Coles}, Rebecca and {Connolly}, Andrew J. and {Cook}, Kem H. and {Cooray}, Asantha and {Covey}, Kevin R. and {Cribbs}, Chris and {Cui}, Wei and {Cutri}, Roc and {Daly}, Philip N. and {Daniel}, Scott F. and {Daruich}, Felipe and {Daubard}, Guillaume and {Daues}, Greg and {Dawson}, William and {Delgado}, Francisco and {Dellapenna}, Alfred and {de Peyster}, Robert and {de Val-Borro}, Miguel and {Digel}, Seth W. and {Doherty}, Peter and {Dubois}, Richard and {Dubois-Felsmann}, Gregory P. and {Durech}, Josef and {Economou}, Frossie and {Eifler}, Tim and {Eracleous}, Michael and {Emmons}, Benjamin L. and {Fausti Neto}, Angelo and {Ferguson}, Henry and {Figueroa}, Enrique and {Fisher-Levine}, Merlin and {Focke}, Warren and {Foss}, Michael D. and {Frank}, James and {Freemon}, Michael D. and {Gangler}, Emmanuel and {Gawiser}, Eric and {Geary}, John C. and {Gee}, Perry and {Geha}, Marla and {Gessner}, Charles J.~B. and {Gibson}, Robert R. and {Gilmore}, D. Kirk and {Glanzman}, Thomas and {Glick}, William and {Goldina}, Tatiana and {Goldstein}, Daniel A. and {Goodenow}, Iain and {Graham}, Melissa L. and {Gressler}, William J. and {Gris}, Philippe and {Guy}, Leanne P. and {Guyonnet}, Augustin and {Haller}, Gunther and {Harris}, Ron and {Hascall}, Patrick A. and {Haupt}, Justine and {Hernandez}, Fabio and {Herrmann}, Sven and {Hileman}, Edward and {Hoblitt}, Joshua and {Hodgson}, John A. and {Hogan}, Craig and {Howard}, James D. and {Huang}, Dajun and {Huffer}, Michael E. and {Ingraham}, Patrick and {Innes}, Walter R. and {Jacoby}, Suzanne H. and {Jain}, Bhuvnesh and {Jammes}, Fabrice and {Jee}, M. James and {Jenness}, Tim and {Jernigan}, Garrett and {Jevremovi{\'c}}, Darko and {Johns}, Kenneth and {Johnson}, Anthony S. and {Johnson}, Margaret W.~G. and {Jones}, R. Lynne and {Juramy-Gilles}, Claire and {Juri{\'c}}, Mario and {Kalirai}, Jason S. and {Kallivayalil}, Nitya J. and {Kalmbach}, Bryce and {Kantor}, Jeffrey P. and {Karst}, Pierre and {Kasliwal}, Mansi M. and {Kelly}, Heather and {Kessler}, Richard and {Kinnison}, Veronica and {Kirkby}, David and {Knox}, Lloyd and {Kotov}, Ivan V. and {Krabbendam}, Victor L. and {Krughoff}, K. Simon and {Kub{\'a}nek}, Petr and {Kuczewski}, John and {Kulkarni}, Shri and {Ku}, John and {Kurita}, Nadine R. and {Lage}, Craig S. and {Lambert}, Ron and {Lange}, Travis and {Langton}, J. Brian and {Le Guillou}, Laurent and {Levine}, Deborah and {Liang}, Ming and {Lim}, Kian-Tat and {Lintott}, Chris J. and {Long}, Kevin E. and {Lopez}, Margaux and {Lotz}, Paul J. and {Lupton}, Robert H. and {Lust}, Nate B. and {MacArthur}, Lauren A. and {Mahabal}, Ashish and {Mandelbaum}, Rachel and {Markiewicz}, Thomas W. and {Marsh}, Darren S. and {Marshall}, Philip J. and {Marshall}, Stuart and {May}, Morgan and {McKercher}, Robert and {McQueen}, Michelle and {Meyers}, Joshua and {Migliore}, Myriam and {Miller}, Michelle and {Mills}, David J.},
        title = "{LSST: From Science Drivers to Reference Design and Anticipated Data Products}",
      journal = {\apj},
         year = 2019,
        month = mar,
       volume = {873},
       number = {2},
          eid = {111},
        pages = {111},
          doi = {10.3847/1538-4357/ab042c},
archivePrefix = {arXiv},
       eprint = {0805.2366},
 primaryClass = {astro-ph},
       adsurl = {https://ui.adsabs.harvard.edu/abs/2019ApJ...873..111I}
}

@ARTICLE{jin2024,
       author = {{Jin}, Shoko and {Trager}, Scott C. and {Dalton}, Gavin B. and {Aguerri}, J. Alfonso L. and {Drew}, J.~E. and {Falc{\'o}n-Barroso}, Jes{\'u}s and {G{\"a}nsicke}, Boris T. and {Hill}, Vanessa and {Iovino}, Angela and {Pieri}, Matthew M. and {Poggianti}, Bianca M. and {Smith}, D.~J.~B. and {Vallenari}, Antonella and {Abrams}, Don Carlos and {Aguado}, David S. and {Antoja}, Teresa and {Arag{\'o}n-Salamanca}, Alfonso and {Ascasibar}, Yago and {Babusiaux}, Carine and {Balcells}, Marc and {Barrena}, R. and {Battaglia}, Giuseppina and {Belokurov}, Vasily and {Bensby}, Thomas and {Bonifacio}, Piercarlo and {Bragaglia}, Angela and {Carrasco}, Esperanza and {Carrera}, Ricardo and {Cornwell}, Daniel J. and {Dom{\'\i}nguez-Palmero}, Lilian and {Duncan}, Kenneth J. and {Famaey}, Benoit and {Fari{\~n}a}, Cecilia and {Gonzalez}, Oscar A. and {Guest}, Steve and {Hatch}, Nina A. and {Hess}, Kelley M. and {Hoskin}, Matthew J. and {Irwin}, Mike and {Knapen}, Johan H. and {Koposov}, Sergey E. and {Kuchner}, Ulrike and {Laigle}, Clotilde and {Lewis}, Jim and {Longhetti}, Marcella and {Lucatello}, Sara and {M{\'e}ndez-Abreu}, Jairo and {Mercurio}, Amata and {Molaeinezhad}, Alireza and {Mongui{\'o}}, Maria and {Morrison}, Sean and {Murphy}, David N.~A. and {Peralta de Arriba}, Luis and {P{\'e}rez}, Isabel and {P{\'e}rez-R{\`a}fols}, Ignasi and {Pic{\'o}}, Sergio and {Raddi}, Roberto and {Romero-G{\'o}mez}, Merc{\`e} and {Royer}, Fr{\'e}d{\'e}ric and {Siebert}, Arnaud and {Seabroke}, George M. and {Som}, Debopam and {Terrett}, David and {Thomas}, Guillaume and {Wesson}, Roger and {Worley}, C. Clare and {Alfaro}, Emilio J. and {Allende Prieto}, Carlos and {Alonso-Santiago}, Javier and {Amos}, Nicholas J. and {Ashley}, Richard P. and {Balaguer-N{\'u}{\~n}ez}, Lola and {Balbinot}, Eduardo and {Bellazzini}, Michele and {Benn}, Chris R. and {Berlanas}, Sara R. and {Bernard}, Edouard J. and {Best}, Philip and {Bettoni}, Daniela and {Bianco}, Andrea and {Bishop}, Georgia and {Blomqvist}, Michael and {Boeche}, Corrado and {Bolzonella}, Micol and {Bonoli}, Silvia and {Bosma}, Albert and {Britavskiy}, Nikolay and {Busarello}, Gianni and {Caffau}, Elisabetta and {Cantat-Gaudin}, Tristan and {Castro-Ginard}, Alfred and {Couto}, Guilherme and {Carbajo-Hijarrubia}, Juan and {Carter}, David and {Casamiquela}, Laia and {Conrado}, Ana M. and {Corcho-Caballero}, Pablo and {Costantin}, Luca and {Deason}, Alis and {de Burgos}, Abel and {De Grandi}, Sabrina and {Di Matteo}, Paola and {Dom{\'\i}nguez-G{\'o}mez}, Jes{\'u}s and {Dorda}, Ricardo and {Drake}, Alyssa and {Dutta}, Rajeshwari and {Erkal}, Denis and {Feltzing}, Sofia and {Ferr{\'e}-Mateu}, Anna and {Feuillet}, Diane and {Figueras}, Francesca and {Fossati}, Matteo and {Franciosini}, Elena and {Frasca}, Antonio and {Fumagalli}, Michele and {Gallazzi}, Anna and {Garc{\'\i}a-Benito}, Rub{\'e}n and {Gentile Fusillo}, Nicola and {Gebran}, Marwan and {Gilbert}, James and {Gledhill}, T.~M. and {Gonz{\'a}lez Delgado}, Rosa M. and {Greimel}, Robert and {Guarcello}, Mario Giuseppe and {Guerra}, Jose and {Gullieuszik}, Marco and {Haines}, Christopher P. and {Hardcastle}, Martin J. and {Harris}, Amy and {Haywood}, Misha and {Helmi}, Amina and {Hernandez}, Nauzet and {Herrero}, Artemio and {Hughes}, Sarah and {Ir{\v{s}}i{\v{c}}}, Vid and {Jablonka}, Pascale and {Jarvis}, Matt J. and {Jordi}, Carme and {Kondapally}, Rohit and {Kordopatis}, Georges and {Krogager}, Jens-Kristian and {La Barbera}, Francesco and {Lam}, Man I. and {Larsen}, S{\o}ren S. and {Lemasle}, Bertrand and {Lewis}, Ian J. and {Lhom{\'e}}, Emilie and {Lind}, Karin and {Lodi}, Marcello and {Longobardi}, Alessia and {Lonoce}, Ilaria and {Magrini}, Laura and {Ma{\'\i}z Apell{\'a}niz}, Jes{\'u}s and {Marchal}, Olivier and {Marco}, Amparo and {Martin}, Nicolas F. and {Matsuno}, Tadafumi and {Maurogordato}, Sophie and {Merluzzi}, Paola and {Miralda-Escud{\'e}}, Jordi and {Molinari}, Emilio and {Monari}, Giacomo and {Morelli}, Lorenzo and {Mottram}, Christopher J. and {Naylor}, Tim and {Negueruela}, Ignacio and {O{\~n}orbe}, Jose and {Pancino}, Elena and {Peirani}, S{\'e}bastien and {Peletier}, Reynier F. and {Pozzetti}, Lucia and {Rainer}, Monica and {Ramos}, Pau and {Read}, Shaun C. and {Rossi}, Elena Maria and {R{\"o}ttgering}, Huub J.~A. and {Rubi{\~n}o-Mart{\'\i}n}, Jose Alberto and {Sabater}, Jose and {San Juan}, Jos{\'e} and {Sanna}, Nicoletta and {Schallig}, Ellen and {Schiavon}, Ricardo P. and {Schultheis}, Mathias and {Serra}, Paolo and {Shimwell}, Timothy W. and {Sim{\'o}n-D{\'\i}az}, Sergio and {Smith}, Russell J. and {Sordo}, Rosanna and {Sorini}, Daniele and {Soubiran}, Caroline and {Starkenburg}, Else and {Steele}, Iain A. and {Stott}, John and {Stuik}, Remko and {Tolstoy}, Eline and {Tortora}, Crescenzo and {Tsantaki}, Maria and {Van der Swaelmen}, Mathieu and {van Weeren}, Reinout J. and {Vergani}, Daniela},
        title = "{The wide-field, multiplexed, spectroscopic facility WEAVE: Survey design, overview, and simulated implementation}",
      journal = {\mnras},
         year = 2024,
        month = may,
       volume = {530},
       number = {3},
        pages = {2688-2730},
          doi = {10.1093/mnras/stad557},
archivePrefix = {arXiv},
       eprint = {2212.03981},
 primaryClass = {astro-ph.IM},
       adsurl = {https://ui.adsabs.harvard.edu/abs/2024MNRAS.530.2688J}
}

@ARTICLE{joo2025,
       author = {{Joo}, Hyungjin and {Jee}, M. James and {Kim}, Juhan and {Lee}, Jaehyun and {Ko}, Jongwan and {Park}, Changbom and {Shin}, Jihye and {Snaith}, Owain and {Pichon}, Christophe and {Gibson}, Brad and {Kim}, Yonghwi},
        title = "{Tracing the Formation History of Intrahalo Light with Horizon Run 5}",
      journal = {\apj},
         year = 2025,
        month = sep,
       volume = {990},
       number = {2},
          eid = {96},
        pages = {96},
          doi = {10.3847/1538-4357/adf4d0},
archivePrefix = {arXiv},
       eprint = {2411.08117},
 primaryClass = {astro-ph.GA},
       adsurl = {https://ui.adsabs.harvard.edu/abs/2025ApJ...990...96J}
}

@ARTICLE{kalirai2006,
       author = {{Kalirai}, Jasonjot S. and {Gilbert}, Karoline M. and {Guhathakurta}, Puragra and {Majewski}, Steven R. and {Ostheimer}, James C. and {Rich}, R. Michael and {Cooper}, Michael C. and {Reitzel}, David B. and {Patterson}, Richard J.},
        title = "{The Metal-poor Halo of the Andromeda Spiral Galaxy (M31)1,}",
      journal = {\apj},
         year = 2006,
        month = sep,
       volume = {648},
       number = {1},
        pages = {389-404},
          doi = {10.1086/505697},
archivePrefix = {arXiv},
       eprint = {astro-ph/0605170},
 primaryClass = {astro-ph},
       adsurl = {https://ui.adsabs.harvard.edu/abs/2006ApJ...648..389K}
}

@ARTICLE{lane2022,
       author = {{Lane}, James M.~M. and {Bovy}, Jo and {Mackereth}, J. Ted},
        title = "{The kinematic properties of Milky Way stellar halo populations}",
      journal = {\mnras},
         year = 2022,
        month = mar,
       volume = {510},
       number = {4},
        pages = {5119-5141},
          doi = {10.1093/mnras/stab3755},
archivePrefix = {arXiv},
       eprint = {2106.09699},
 primaryClass = {astro-ph.GA},
       adsurl = {https://ui.adsabs.harvard.edu/abs/2022MNRAS.510.5119L}
}

@ARTICLE{lee2017,
       author = {{Lee}, Jaehyun and {Yi}, Sukyoung K.},
        title = "{Formation and Assembly History of Stellar Components in Galaxies as a Function of Stellar and Halo Mass}",
      journal = {\apj},
         year = 2017,
        month = feb,
       volume = {836},
       number = {2},
          eid = {161},
        pages = {161},
          doi = {10.3847/1538-4357/aa5b87},
archivePrefix = {arXiv},
       eprint = {1701.07527},
 primaryClass = {astro-ph.GA},
       adsurl = {https://ui.adsabs.harvard.edu/abs/2017ApJ...836..161L}
}

@ARTICLE{merritt2016,
       author = {{Merritt}, Allison and {van Dokkum}, Pieter and {Abraham}, Roberto and {Zhang}, Jielai},
        title = "{The Dragonfly nearby Galaxies Survey. I. Substantial Variation in the Diffuse Stellar Halos around Spiral Galaxies}",
      journal = {\apj},
         year = 2016,
        month = oct,
       volume = {830},
       number = {2},
          eid = {62},
        pages = {62},
          doi = {10.3847/0004-637X/830/2/62},
archivePrefix = {arXiv},
       eprint = {1606.08847},
 primaryClass = {astro-ph.GA},
       adsurl = {https://ui.adsabs.harvard.edu/abs/2016ApJ...830...62M}
}

@ARTICLE{mihos2017,
       author = {{Mihos}, J. Christopher and {Harding}, Paul and {Feldmeier}, John J. and {Rudick}, Craig and {Janowiecki}, Steven and {Morrison}, Heather and {Slater}, Colin and {Watkins}, Aaron},
        title = "{The Burrell Schmidt Deep Virgo Survey: Tidal Debris, Galaxy Halos, and Diffuse Intracluster Light in the Virgo Cluster}",
      journal = {\apj},
         year = 2017,
        month = jan,
       volume = {834},
       number = {1},
          eid = {16},
        pages = {16},
          doi = {10.3847/1538-4357/834/1/16},
archivePrefix = {arXiv},
       eprint = {1611.04435},
 primaryClass = {astro-ph.GA},
       adsurl = {https://ui.adsabs.harvard.edu/abs/2017ApJ...834...16M}
}

@ARTICLE{monachesi2016,
       author = {{Monachesi}, Antonela and {G{\'o}mez}, Facundo A. and {Grand}, Robert J.~J. and {Kauffmann}, Guinevere and {Marinacci}, Federico and {Pakmor}, R{\"u}diger and {Springel}, Volker and {Frenk}, Carlos S.},
        title = "{On the stellar halo metallicity profile of Milky Way-like galaxies in the Auriga simulations}",
      journal = {\mnras},
         year = 2016,
        month = jun,
       volume = {459},
       number = {1},
        pages = {L46-L50},
          doi = {10.1093/mnrasl/slw052},
archivePrefix = {arXiv},
       eprint = {1512.03064},
 primaryClass = {astro-ph.GA},
       adsurl = {https://ui.adsabs.harvard.edu/abs/2016MNRAS.459L..46M}
}

@ARTICLE{monachesi2019,
       author = {{Monachesi}, Antonela and {G{\'o}mez}, Facundo A. and {Grand}, Robert J.~J. and {Simpson}, Christine M. and {Kauffmann}, Guinevere and {Bustamante}, Sebasti{\'a}n and {Marinacci}, Federico and {Pakmor}, R{\"u}diger and {Springel}, Volker and {Frenk}, Carlos S. and {White}, Simon D.~M. and {Tissera}, Patricia B.},
        title = "{The Auriga stellar haloes: connecting stellar population properties with accretion and merging history}",
      journal = {\mnras},
         year = 2019,
        month = may,
       volume = {485},
       number = {2},
        pages = {2589-2616},
          doi = {10.1093/mnras/stz538},
archivePrefix = {arXiv},
       eprint = {1804.07798},
 primaryClass = {astro-ph.GA},
       adsurl = {https://ui.adsabs.harvard.edu/abs/2019MNRAS.485.2589M}
}

@ARTICLE{montenegro-taborda2023,
       author = {{Montenegro-Taborda}, Daniel and {Rodriguez-Gomez}, Vicente and {Pillepich}, Annalisa and {Avila-Reese}, Vladimir and {Sales}, Laura V. and {Rodr{\'\i}guez-Puebla}, Aldo and {Hernquist}, Lars},
        title = "{The growth of brightest cluster galaxies in the TNG300 simulation: dissecting the contributions from mergers and in situ star formation}",
      journal = {\mnras},
         year = 2023,
        month = may,
       volume = {521},
       number = {1},
        pages = {800-817},
          doi = {10.1093/mnras/stad586},
archivePrefix = {arXiv},
       eprint = {2302.10943},
 primaryClass = {astro-ph.GA},
       adsurl = {https://ui.adsabs.harvard.edu/abs/2023MNRAS.521..800M}
}

@ARTICLE{montes2018,
       author = {{Montes}, Mireia and {Trujillo}, Ignacio},
        title = "{Intracluster light at the Frontier - II. The Frontier Fields Clusters}",
      journal = {\mnras},
         year = 2018,
        month = feb,
       volume = {474},
       number = {1},
        pages = {917-932},
          doi = {10.1093/mnras/stx2847},
archivePrefix = {arXiv},
       eprint = {1710.03240},
 primaryClass = {astro-ph.CO},
       adsurl = {https://ui.adsabs.harvard.edu/abs/2018MNRAS.474..917M}
}

@ARTICLE{mouhcine2005,
       author = {{Mouhcine}, M. and {Ferguson}, H.~C. and {Rich}, R.~M. and {Brown}, T.~M. and {Smith}, T.~E.},
        title = "{Halos of Spiral Galaxies. I. The Tip of the Red Giant Branch as a Distance Indicator}",
      journal = {\apj},
         year = 2005,
        month = nov,
       volume = {633},
       number = {2},
        pages = {810-820},
          doi = {10.1086/468177},
archivePrefix = {arXiv},
       eprint = {astro-ph/0510253},
 primaryClass = {astro-ph},
       adsurl = {https://ui.adsabs.harvard.edu/abs/2005ApJ...633..810M}
}

@ARTICLE{nelson2018,
       author = {{Nelson}, Dylan and {Pillepich}, Annalisa and {Springel}, Volker and {Weinberger}, Rainer and {Hernquist}, Lars and {Pakmor}, R{\"u}diger and {Genel}, Shy and {Torrey}, Paul and {Vogelsberger}, Mark and {Kauffmann}, Guinevere and {Marinacci}, Federico and {Naiman}, Jill},
        title = "{First results from the IllustrisTNG simulations: the galaxy colour bimodality}",
      journal = {\mnras},
         year = 2018,
        month = mar,
       volume = {475},
       number = {1},
        pages = {624-647},
          doi = {10.1093/mnras/stx3040},
archivePrefix = {arXiv},
       eprint = {1707.03395},
 primaryClass = {astro-ph.GA},
       adsurl = {https://ui.adsabs.harvard.edu/abs/2018MNRAS.475..624N}
}

@ARTICLE{nelson2019,
       author = {{Nelson}, Dylan and {Springel}, Volker and {Pillepich}, Annalisa and {Rodriguez-Gomez}, Vicente and {Torrey}, Paul and {Genel}, Shy and {Vogelsberger}, Mark and {Pakmor}, Ruediger and {Marinacci}, Federico and {Weinberger}, Rainer and {Kelley}, Luke and {Lovell}, Mark and {Diemer}, Benedikt and {Hernquist}, Lars},
        title = "{The IllustrisTNG simulations: public data release}",
      journal = {Computational Astrophysics and Cosmology},
         year = 2019,
        month = may,
       volume = {6},
       number = {1},
          eid = {2},
        pages = {2},
          doi = {10.1186/s40668-019-0028-x},
archivePrefix = {arXiv},
       eprint = {1812.05609},
 primaryClass = {astro-ph.GA},
       adsurl = {https://ui.adsabs.harvard.edu/abs/2019ComAC...6....2N}
}

@ARTICLE{nelson2024,
       author = {{Nelson}, Dylan and {Pillepich}, Annalisa and {Ayromlou}, Mohammadreza and {Lee}, Wonki and {Lehle}, Katrin and {Rohr}, Eric and {Truong}, Nhut},
        title = "{Introducing the TNG-Cluster simulation: Overview and the physical properties of the gaseous intracluster medium}",
      journal = {\aap},
         year = 2024,
        month = jun,
       volume = {686},
          eid = {A157},
        pages = {A157},
          doi = {10.1051/0004-6361/202348608},
archivePrefix = {arXiv},
       eprint = {2311.06338},
 primaryClass = {astro-ph.GA},
       adsurl = {https://ui.adsabs.harvard.edu/abs/2024A&A...686A.157N}
}

@ARTICLE{nfw1997,
       author = {{Navarro}, Julio F. and {Frenk}, Carlos S. and {White}, Simon D.~M.},
        title = "{A Universal Density Profile from Hierarchical Clustering}",
      journal = {\apj},
         year = 1997,
        month = dec,
       volume = {490},
       number = {2},
        pages = {493-508},
          doi = {10.1086/304888},
archivePrefix = {arXiv},
       eprint = {astro-ph/9611107},
 primaryClass = {astro-ph},
       adsurl = {https://ui.adsabs.harvard.edu/abs/1997ApJ...490..493N}
}

@ARTICLE{oliva-altamirano2014,
       author = {{Oliva-Altamirano}, P. and {Brough}, S. and {Lidman}, C. and {Couch}, W.~J. and {Hopkins}, A.~M. and {Colless}, M. and {Taylor}, E. and {Robotham}, A.~S.~G. and {Gunawardhana}, M.~L.~P. and {Ponman}, T. and {Baldry}, I. and {Bauer}, A.~E. and {Bland-Hawthorn}, J. and {Cluver}, M. and {Cameron}, E. and {Conselice}, C.~J. and {Driver}, S. and {Edge}, A.~C. and {Graham}, A.~W. and {van Kampen}, E. and {Lara-L{\'o}pez}, M.~A. and {Liske}, J. and {L{\'o}pez-S{\'a}nchez}, A.~R. and {Loveday}, J. and {Mahajan}, S. and {Peacock}, J. and {Phillipps}, S. and {Pimbblet}, K.~A. and {Sharp}, R.~G.},
        title = "{Galaxy And Mass Assembly (GAMA): testing galaxy formation models through the most massive galaxies in the Universe}",
      journal = {\mnras},
         year = 2014,
        month = may,
       volume = {440},
       number = {1},
        pages = {762-775},
          doi = {10.1093/mnras/stu277},
archivePrefix = {arXiv},
       eprint = {1402.4139},
 primaryClass = {astro-ph.CO},
       adsurl = {https://ui.adsabs.harvard.edu/abs/2014MNRAS.440..762O}
}

@ARTICLE{pinna2024,
       author = {{Pinna}, Francesca and {Walo-Mart{\'\i}n}, Daniel and {Grand}, Robert J.~J. and {Martig}, Marie and {Fragkoudi}, Francesca and {G{\'o}mez}, Facundo A. and {Marinacci}, Federico and {Pakmor}, R{\"u}diger},
        title = "{Stellar populations and the origin of thick disks in AURIGA simulations}",
      journal = {\aap},
         year = 2024,
        month = mar,
       volume = {683},
          eid = {A236},
        pages = {A236},
          doi = {10.1051/0004-6361/202347388},
archivePrefix = {arXiv},
       eprint = {2311.13700},
 primaryClass = {astro-ph.GA},
       adsurl = {https://ui.adsabs.harvard.edu/abs/2024A&A...683A.236P}
}

@ARTICLE{pillepich2018,
       author = {{Pillepich}, Annalisa and {Nelson}, Dylan and {Hernquist}, Lars and {Springel}, Volker and {Pakmor}, R{\"u}diger and {Torrey}, Paul and {Weinberger}, Rainer and {Genel}, Shy and {Naiman}, Jill P. and {Marinacci}, Federico and {Vogelsberger}, Mark},
        title = "{First results from the IllustrisTNG simulations: the stellar mass content of groups and clusters of galaxies}",
      journal = {\mnras},
         year = 2018,
        month = mar,
       volume = {475},
       number = {1},
        pages = {648-675},
          doi = {10.1093/mnras/stx3112},
archivePrefix = {arXiv},
       eprint = {1707.03406},
 primaryClass = {astro-ph.GA},
       adsurl = {https://ui.adsabs.harvard.edu/abs/2018MNRAS.475..648P}
}

@ARTICLE{planck2020,
       author = {{Planck Collaboration} and {Aghanim}, N. and {Akrami}, Y. and {Ashdown}, M. and {Aumont}, J. and {Baccigalupi}, C. and {Ballardini}, et al.},
        title = "{Planck 2018 results. VI. Cosmological parameters}",
      journal = {\aap},
         year = 2020,
        month = sep,
       volume = {641},
          eid = {A6},
        pages = {A6},
          doi = {10.1051/0004-6361/201833910},
archivePrefix = {arXiv},
       eprint = {1807.06209},
 primaryClass = {astro-ph.CO},
       adsurl = {https://ui.adsabs.harvard.edu/abs/2020A&A...641A...6P}
}

@ARTICLE{prada2012,
       author = {{Prada}, Francisco and {Klypin}, Anatoly A. and {Cuesta}, Antonio J. and {Betancort-Rijo}, Juan E. and {Primack}, Joel},
        title = "{Halo concentrations in the standard {\ensuremath{\Lambda}} cold dark matter cosmology}",
      journal = {\mnras},
         year = 2012,
        month = jul,
       volume = {423},
       number = {4},
        pages = {3018-3030},
          doi = {10.1111/j.1365-2966.2012.21007.x},
archivePrefix = {arXiv},
       eprint = {1104.5130},
 primaryClass = {astro-ph.CO},
       adsurl = {https://ui.adsabs.harvard.edu/abs/2012MNRAS.423.3018P}
}

\appendix
\section{Resolution tests}
\label{sec:app_resolution}

We assess the impact of numerical resolution on the main scaling relations discussed in this work by comparing results obtained from simulations with different mass resolutions over the stellar-mass ranges where they overlap. In particular, we analyse the relations between SH mass and transition radius, SH mass and BCG stellar mass, and SH mass and ICL mass.

\begin{figure}[t!]
\centering
\includegraphics[width=0.48\textwidth]{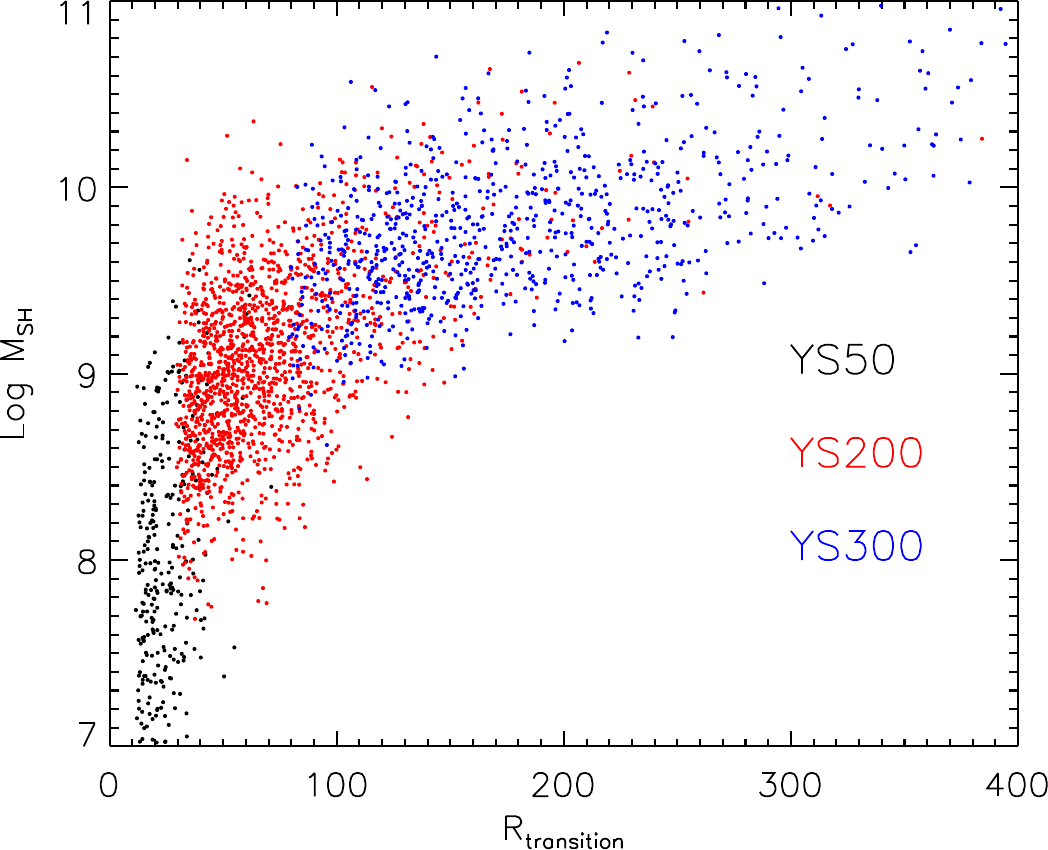}
\caption{Relation between SH mass and transition radius $R_{\rm trans}$ for simulations with different mass resolutions: YS50 (black), YS200 (red), and YS300 (blue). The relations are shown over the stellar-mass ranges covered by each of them. The median trends are consistent within the scatter, with no systematic dependence on numerical resolution, indicating that the SH--$R_{\rm trans}$ relation is robust against resolution effects.}
\label{fig:A1}
\end{figure}

\begin{figure}[t!]
\centering
\includegraphics[width=0.47\textwidth]{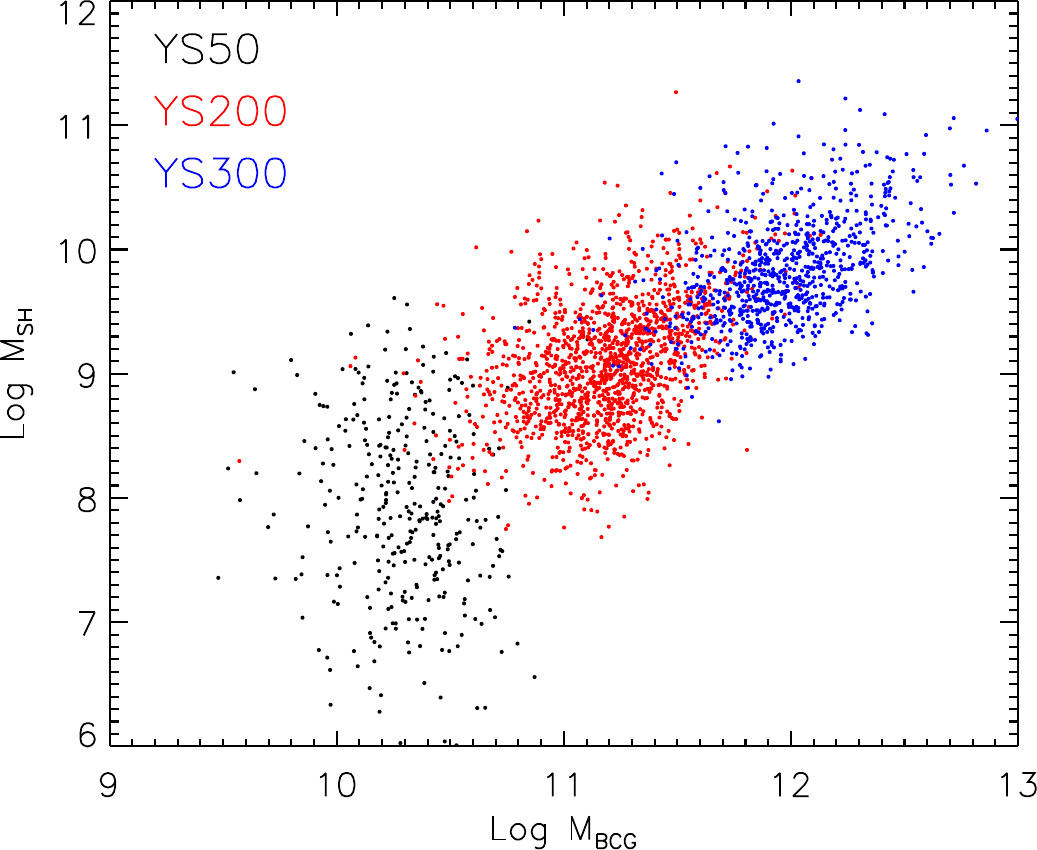}
\\
\includegraphics[width=0.47\textwidth]{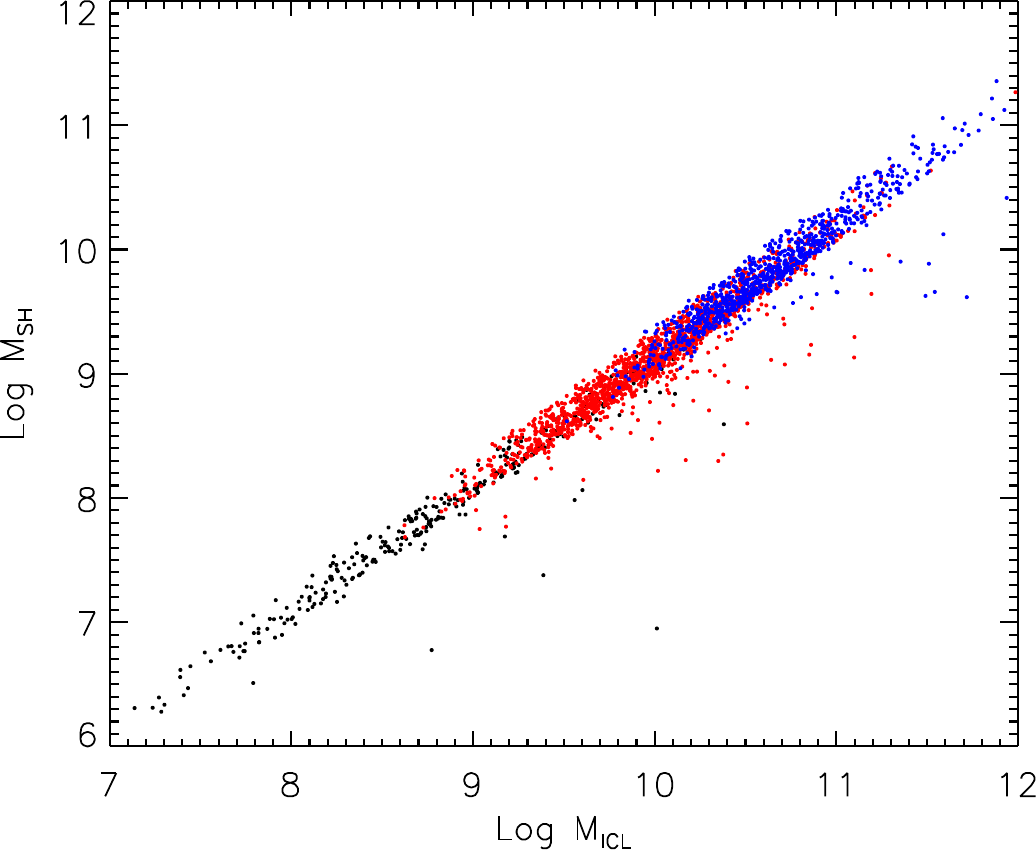}
\caption{Scaling relations between SH mass and BCG stellar mass (upper panel), and SH mass and ICL mass (bottom panel), for simulations with different mass resolutions: YS50 (black), YS200 (red), and YS300 (blue). The agreement between the median relations in the overlapping mass ranges demonstrates that the main scaling relations discussed in the paper are not driven by numerical resolution.}
\label{fig:A2}
\end{figure}

Figure~\ref{fig:A1} shows the relation between SH mass and $R_{\rm trans}$ for the different simulations. The median trends and scatter are consistent within the overlapping mass ranges, with no systematic offsets as a function of resolution. Similarly, Figure~\ref{fig:A2} compares the SH--BCG and SH--ICL mass relations, which also display fully consistent behaviour across simulations of different resolution.

These tests indicate that the scaling relations presented in the main text are not driven by numerical resolution effects. This result is expected, as the semi-analytic model traces integrated stellar components whose global properties primarily depend on halo assembly histories rather than on the detailed resolution of individual substructures.

We therefore conclude that numerical resolution does not significantly affect the trends discussed in this work. For this reason, additional properties such as colours and metallicities, which are derived from the same integrated stellar populations, are not expected to show a stronger dependence on resolution.

\end{document}